\documentclass[prl,preprint,
superscriptaddress,
amsmath,amssymb,
aps,
]{revtex4-2}

\usepackage{graphicx}
\usepackage{dcolumn}
\usepackage{bm}
\usepackage[dvipsnames]{xcolor}
\usepackage[hidelinks]{hyperref}
\usepackage{physics}
\usepackage[normalem]{ulem}

\newcommand{\BiSe}{\mathrm{Bi}_2\mathrm{Se}_3}

\newenvironment{supp}{\color{black}}{}

\begin{document}

\title{Resolving the phase of a Dirac topological state via interferometric photoemission}

\author{Shiri Gvishi}
\thanks{These authors contributed equally to this work.}
\author{Ittai Sidilkover}
\thanks{These authors contributed equally to this work.}
\affiliation{School of Physics and Astronomy, Faculty of Exact Sciences, Tel Aviv University, Tel-Aviv, 6997801, Israel}
\affiliation{Center for Light-Matter Interaction, Tel Aviv University, Tel-Aviv, 6997801, Israel}
\author{Yun Yen}
\thanks{These authors contributed equally to this work.}
\affiliation{Institute for Theoretical Physics and Bremen Center for Computational Materials Science, University of Bremen, 28359 Bremen, Germany}
\author{Shaked Rosenstein}
\author{Nir Hen Levin}
\author{Adi Perelmuter}
\author{Omer Pasternak}
\affiliation{School of Physics and Astronomy, Faculty of Exact Sciences, Tel Aviv University, Tel-Aviv, 6997801, Israel}
\affiliation{Center for Light-Matter Interaction, Tel Aviv University, Tel-Aviv, 6997801, Israel}

\author{Costel R. Rotundu}
\affiliation{Stanford Institute for Materials and Energy Sciences, SLAC National Accelerator Laboratory, 2575 Sand Hill Road, Menlo Park, California 94025, USA}
\affiliation{Geballe Laboratory for Advanced Materials, Stanford University, Stanford, California 94305, USA}

\author{Ido Biran}
\author{Semën Gorfman}
\affiliation{Department of Materials Science and Engineering, Tel Aviv University, Tel Aviv, Israel}

\author{Naaman Amer}
\affiliation{School of Physics and Astronomy, Faculty of Exact Sciences, Tel Aviv University, Tel-Aviv, 6997801, Israel}

\author{Michael A. Sentef}
\email{michael.sentef@mpsd.mpg.de}
\affiliation{Institute for Theoretical Physics and Bremen Center for Computational Materials Science, University of Bremen, 28359 Bremen, Germany}
\affiliation{Max Planck Institute for the Structure and Dynamics of Matter, Center for Free-Electron Laser Science (CFEL), Luruper Chaussee 149, 22761 Hamburg, Germany}

\author{Hadas Soifer}
\email{hadassoifer@tauex.tau.ac.il}
\affiliation{School of Physics and Astronomy, Faculty of Exact Sciences, Tel Aviv University, Tel-Aviv, 6997801, Israel}
\affiliation{Center for Light-Matter Interaction, Tel Aviv University, Tel-Aviv, 6997801, Israel}

\begin{abstract} 
The electronic wavefunction is at the heart of physical phenomena, defining the frontiers of quantum materials research. While the amplitude of the electron wavefunction in crystals can be measured with state-of-the-art probes in unprecedented resolution, its phase has remained largely inaccessible, obscuring rich electronic information. Here we develop a quantum-path electron interferometer based on time- and angle-resolved photoemission spectroscopy, that enables the reconstruction of phase information associated with electronic states, as encoded in the photoemission transition amplitudes
-- with energy and momentum resolution. We demonstrate the scheme by resolving the phase along the Dirac electronic band of a prototypical topological insulator and observe a resonance-associated phase jump as well as a momentum and phase synchronized inversion revealing the helicity of the Dirac cone. We show the interferometer can be optically controlled by the polarization of the absorbed light, allowing a differential measurement of the phase -- a crucial component for extracting phase information from an interferogram.
This photo-electron-interferometer provides direct experimental access to the phase of electronic transition amplitudes. Its implementation relies on experimentally accessible conditions -- such as the presence of a suitable intermediate state and polarization-selective coupling -- and can therefore be extended to a wide class of materials.

\end{abstract}

\maketitle

Retrieving the phase information of a quantum wave-function has been a long-standing challenge since the early days of quantum mechanics. While the phase of light has long been accessible through interferometry, unveiling the wave nature of photons and driving revolutions in optics, physics, and technology, its role in matter has remained far more elusive. Yet it is the complex structure and phase of the many-body wavefunction -- manifested in observable transition amplitudes -- that encodes the internal coherence, correlations, and topology that define the behavior of quantum materials\cite{keimer_physics_2017}. From the Michelson–Morley experiment\cite{Michelson1887On}, through electron double-slit interference\cite{thomson_diffraction_1927},  attosecond interferometry\cite{azoury_electronic_2019}, and gravitational-wave detection\cite{bailes_gravitational-wave_2021}, the pursuit of phase has continually redefined the limits of measurement. Today, the frontier lies in retrieving the phase of electronic states within quantum materials, particularly near singularities such as the Dirac point, where the quantum geometry of the wavefunction gives rise to nontrivial topology and exotic responses\cite{yan_topological_2012}. Conventional probes such as STM (scanning tunneling microscopy\cite{RevModPhys.59.615}) and ARPES (angle-resolved photoemission spectroscopy\cite{Damascelli_2004,lv_angle-resolved_2019}) provide access to amplitude information, revealing where electrons reside and how they move, but leave the phase, and with it the full complexity of the quantum state, experimentally unresolved. Revealing this hidden phase is therefore essential for capturing the full complexity of quantum matter and for uncovering the mechanisms by which topology, symmetry breaking, and coherence interplay at the most fundamental level.

Phase and temporal dynamics are inherently linked: the evolution of a quantum state is encoded in its time-dependent phase, and coherent light-matter interactions can be harnessed to resolve the electronic wave.
Time- and angle-resolved photoemission spectroscopy (trARPES) is a natural candidate to access light-matter interactions in quantum materials. It incorporates the ultrafast pump-probe scheme -- providing femtosecond temporal resolution -- with ARPES -- a cornerstone technique in quantum material research that maps the band structure of crystalline solids by photoemission\cite{boschini_time-resolved_2024,gedik_photoemission_2017}. However, despite its ability to access energy-, momentum-, and time-resolved dynamics, trARPES typically resolves only incoherent population dynamics, where the phase information of the electronic wavefunction still remains inaccessible.  

Retrieving this electronic phase requires an interferometric approach. However, unlike optical fields, constructing an interferometer for matter waves with full energy and momentum resolution poses extreme challenges, particularly under the cryogenic, ultrahigh-vacuum, and high-resolution conditions required for quantum material studies.
Instead of constructing a conventional interferometer where spatially separated beams interfere, we exploit
quantum-path (QP) interference between indistinguishable excitation pathways - to construct a nonlinear interferometer for electronic wavefunctions. Under suitable excitation conditions, two-photon photoemission (2PPE), or trARPES in its nonlinear regime, naturally supports multiple absorption pathways that can interfere coherently, analogous to other nonlinear-optical processes. Although QP interference has been observed in earlier trARPES and 2PPE studies\cite{eickhoff_two-state_2011,chan_communication_2011}, including Floquet-Volkov-type processes\cite{mahmood_selective_2016, bao_floquet-volkov_2025, choi_observation_2025, merboldt_observation_2025, fragkos_floquet-bloch_2025}, and coherent trARPES has provided access to electronic coherence and dephasing\cite{petek_photoexcitation_2012,tao_direct_2016, reutzel_coherent_2020, reutzel_probing_2023}, the momentum-dependent phase encoded in electronic transition amplitudes has remained inaccessible, primarily due to the lack of a controllable phase calibration.

In this manuscript, we introduce an interferometric scheme implemented in trARPES that enables the reconstruction of the phase encoded in electronic transition amplitudes in quantum materials with simultaneous energy and momentum resolution. We implement the scheme in $\BiSe$, a prototypical topological insulator\cite{xia_observation_2009,chen_experimental_2009}, exploiting its strong spin-orbit coupling and polarization-dependent dichroism\cite{ketterl_origin_2018,zhang_probing_2021} to identify and isolate the interference contribution. We focus on the second Dirac cone -- a topological surface state in the unoccupied bands\cite{sobota_direct_2013, soifer_band_2019} -- 
and reconstruct its energy- and momentum-dependent phase, revealing a resonance-associated phase evolution and a momentum-dependent phase inversion linked to the helicity of the Dirac cone.

\begin{figure}
\centering \includegraphics[width=\textwidth]{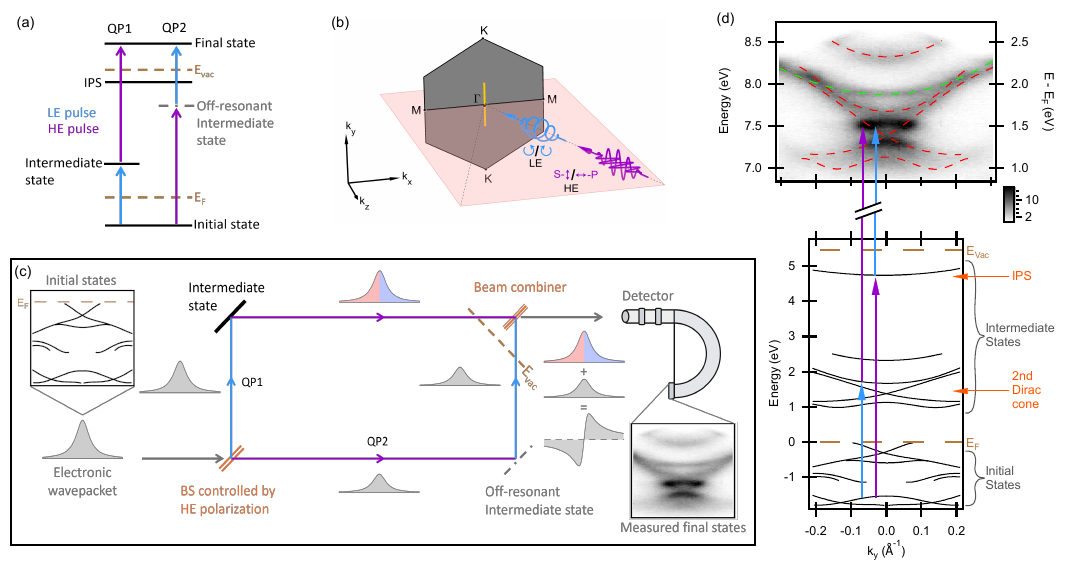}
\vspace{-15mm}
\caption{\textbf{Quantum-path interferometer in trARPES.} LE and HE photons are depicted by the blue and purple arrows, respectively, throughout the figure. (a) Schematic diagram of the two QPs available in two-color 2PPE with the same initial and final states, where QP1 goes through a resonant intermediate state, while QP2 goes through a off-resonant one. (b) Experimental geometry: the plane of incidence is along the $\Gamma-M$ direction (red plane) in the hexagonal Brillouin-zone of $\BiSe$, and the measured momentum cut is along $\Gamma-K$ (yellow line). (c) Conceptual QP-interferometer: The electronic wavefunction originating from the initial states (gray Lorentzian on the left) is split into the two QPs (depicted by the two gray wavepackets in QP1 and QP2). In QP1, the electron wavepacket acquires energy- and momentum-dependent phase due to the resonant transition into the intermediate state, represented by the red-blue shading, while in QP2, the non-resonant transition does not modify the phase structure. Upon absorption of the second photon, both paths reach the same final state (beam-combiner) and the wavepackets interfere. The intensity of the combined wavepacket in energy and momentum is the trARPES spectrum measured on the detector.  (d) Schematic visualization of the optical transitions of both QPs in $\BiSe$. The initial and intermediate states involved in the process (bottom panel) are extracted from the experimental data (see Methods). The top panel presents the measured spectrum of the final states originating from both QPs. The intermediate states of QP1 (QP2), upshifted by $\hbar\omega_\mathrm{HE}$ ($\hbar\omega_\mathrm{LE}$) are marked in dashed red (green). The right axis on the top panel is the energy of QP1 intermediate states relative to the Fermi level.}
\label{fig:1}
\end{figure}

\subsection{Concept of controllable quantum-path interferometer in two-photon photoemission}
We start by introducing the concept of controllable QP interference in a 2PPE process excited by two pulses of different photon energies: one low-photon-energy (LE) pulse and one high-photon-energy (HE) pulse.   
Generally, in a two-color two-photon absorption process, there are two possible pathways, depending on the order of photon absorption (see Fig.~\ref{fig:1}(a)). 
The first quantum-path (QP1) we consider is the one commonly observed in trARPES experiments: the LE pulse excites an electron from a populated initial state to an intermediate (unoccupied) state, and the HE pulse further excites this electron to a final state, leading to photoemission. The second quantum-path (QP2) is the opposite excitation pathway: the HE pulse excites an electron to a high-energy intermediate state, and the LE pulse consequently photoemits the electron from this highly excited state to a final state. This second pathway is typically absent in trARPES, and is only "active" in specific conditions that will be discussed below. 

Temporally, both pathways can occur when the two pulses overlap in time. 
Energetically, when exciting from the same initial state, the two pathways will lead to the same final electron (kinetic) energy, as long as the electron has not scattered between the first and second excitation steps. 
Therefore, the two QPs populate final photoelectron states with the same energy and momentum (Fig.~\ref{fig:1}(a)) where they interfere. Since the QPs start and end at the same electronic energy, the phase difference between them is determined mainly by the intermediate states.
Because each pathway absorbs one photon from each pulse, any pulse-to-pulse phase jitter is common to both paths and cancels. The interference therefore does not originate from a controlled optical phase relationship between the two laser fields, but from the coherent addition of two indistinguishable electronic excitation pathways. The resolved phase is thus electronic in origin and reflects the properties of the quantum pathways inside the material rather than the phase of the driving optical fields.

The observation of an interference signature in itself is not enough for resolving phase. For phase reconstruction, it is necessary to have some degree of control, such as changing the length of one of the interferometer's arms. 
Such control can be hard to achieve with a QP electron interferometer as opposed to a spatial optical interferometer.
In order to achieve the required controllability in a trARPES setup, we aim for the interferometer scheme outlined in Fig.~\ref{fig:1}(c), where the phase of interest is that of QP1, while QP2 serves as a reference arm. In QP1, an intermediate state is resonantly excited by the LE pulse, before photoemission by the HE pulse. For the reference arm we have a few requirements: its phase should have as little structure as possible (in energy and momentum), and it should therefore be a non-resonant 2PPE pathway, as in Fig.~\ref{fig:1}(a). However, in order for this non-resonant transition to have a non-negligible amplitude, it needs to be energetically close to some bound state. Here we will use the polarization selection-rules of such a close-by bound state in order to turn ON and OFF the interfering arm, therefore providing all the information needed for phase reconstruction. 

\subsection{Quantum-path interferometer in a topological insulator}

We implement this electron-QP-interferometer scheme in $\BiSe$, with LE and HE pulses of 3 eV and 6 eV photon energy, respectively, in the experimental geometry as described in Fig.~\ref{fig:1}(b), where $k_y$ is the measured momentum direction, along $\Gamma-K$ (see Methods for further details). 

The spectrum of QP1 is the standard trARPES pump-probe spectrum\cite{sobota_ultrafast_2012,sobota_direct_2013,ketterl_origin_2018,soifer_band_2019}, with the range of relevant initial and intermediate states sketched in Fig.~\ref{fig:1}(d) bottom. 
The second path is often absent in trARPES, both because of pulse-energy considerations (see Methods), and because it requires the presence of high-energy bound states close to the vacuum level, where the band density is typically lower. In $\BiSe$, such a high-lying state exists about 4.8 eV above the Dirac point (0.63 eV below the vacuum level, see arrow in panel (d)). This is the Image Potential State (IPS), which is a generic surface state typical to metals\cite{dose_image_1987,aeschlimann_time-resolved_2025}, and has been well characterized in $\BiSe$\cite{sobota_ultrafast_2012,niesner_electron_2014,bugini_ultrafast_2017}. 
Therefore, QP2 involves excitation with a HE photon into or close to the IPS, and photoemission via a LE photon (Fig.~\ref{fig:1}(d)). The experimental signature of the second path -- when it does go through a resonance -- is a time-reversed trARPES signal which can overlap on the detector with much lower-lying states (Fig.~\ref{fig:1}(d) top). The overlap on the detector of these two paths is well known\cite{reutzel_coherent_2020,biedermann_spin-split_2012,bugini_ultrafast_2017}; however, it is not always appreciated that it is not just the signal intensity summing up on the detector, but a coherent sum of the two signals, which can then lead to interference -- even though the intermediate states are at very different energies. 

Importantly, the coupling to the IPS in $\BiSe$ is strongly dependent on polarization\cite{wolf_direct_1999}: with S-polarized excitation there is close to zero coupling to the state, whereas with P-polarization the coupling is strong (see Methods). This strong polarization dependence provides us with an ON/OFF switch for the interference: when the HE beam is S-polarized, there is no (or very weak) coupling to the IPS, QP2 is disabled, and only the intensity of QP1 is measured on the detector. With a P-polarized HE beam, there is a strong coupling to this state, QP2 is activated, and the signal on the detector comprises of the resulting intensity of the two interfering paths (Fig.~\ref{fig:1}(d) top).
It is important to note that even when the HE excitation is not resonant with the transition into the IPS, the coupling to the IPS still significantly increases the probability of this two-photon transition, and therefore the second path is active in P-polarization and inactive in S, even in far-from-resonance scenarios.  

This system can therefore be used as an electronic interferometer as presented schematically in Fig.~\ref{fig:1}(c), where the input state is an electron in an occupied band. This electronic wavefunction is split into two paths by the optical excitation: LE excitation to one arm, and HE to the other, with the polarization of HE determining the splitting ratio between the two paths. The electron on the HE side (QP2) undergoes a non-resonant transition, providing the reference phase. On the LE side (QP1), it crosses a resonant state such that the phase of this state is imprinted on the electronic wavefunction. Then absorption of a HE photon brings it to the same electronic final state energy as the reference arm, imprinting the interference between the two arms on the intensity that is finally measured on the detector. 
In analogy to optical imaging interferometers, which can spatially resolve the phase for different image locations, here we resolve the electronic phase in energy-momentum space.

\subsection{Identifying interference in trARPES data}

We begin by characterizing our energy-momentum resolved interferometer. The spectrum presented in Fig.~\ref{fig:1}(d) top, includes contributions from QP1 -- red dashed lines -- as well as contributions from QP2 -- green dashed line, however it is unclear how far from its (IPS) resonance QP2 could influence the signal. Moreover, we expect the interference to be visible only in energy-momentum regions where QP1 is resonant, since other non-resonantly excited states, i.e. states into which electrons have scattered, will not maintain the electron's initial phase and will not contribute to the interference. Ideally, the subtraction of the reference spectrum -- where QP2 is turned off by using S-polarized HE pulse -- would highlight all the regions that are affected by the interference. However, one needs to be more careful due to additional matrix-element effects in photoemission: when using S- vs P- polarization in photoemission, different bands may be highlighted depending on their orbital character\cite{gierz_illuminating_2011,moser_toy_2023,sidilkover_reexamining_2025}. 

\begin{figure}
\centering \includegraphics[width=\textwidth]{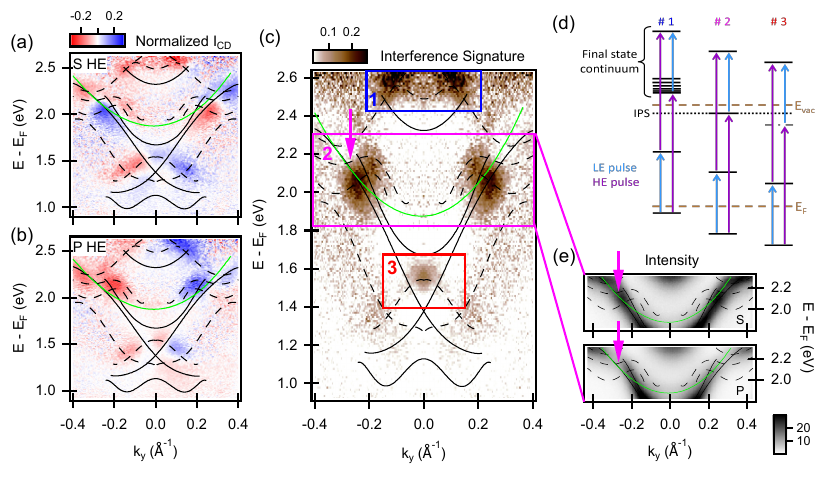}
\vspace{-18mm}
\caption{\textbf{Identifying interference regions}. 
(a-b) Normalized $I_\mathrm{CD}$ spectra for S- and P-polarized HE, respectively. (c) k-symmetrized difference between the CD spectra in (a) and (b), showing the interference signatures. Regions displaying significant interference are highlighted by the numbered rectangles. (d) Schematic 2PPE pathway pairs (QP1 - left, QP2 - right), corresponding to the spectral regions \#1-3 in (c). (e) 2PPE intensity ($I^-+I^+$) of region \#2 for S-pol HE (top) and P-pol HE (bottom). Magenta arrows highlight the region where destructive interference is observed in (e).  For all spectra in the figure, the intermediate states for QP1 (QP2) are marked by solid black (green) lines. The upshifted initial states (by LE = 3.08 eV) are marked by black dashed lines. 
}
\label{fig:Interf_vs_matE}
\end{figure}

In order to identify where the S- vs P- polarization difference goes beyond just matrix-element effect and indicates a contribution from a second photoemission channel, we take advantage of circular-dichroism (CD) of the LE-pulse (i.e difference in photoemission intensity measured with $\sigma_+$ and $\sigma_-$ polarized LE).
We first analyze the signal under the assumption of \textbf{one channel contribution} (QP1 only), and show that our measurements do not agree with the one-photoemission-channel interpretation. Using second-order perturbation analysis (see Supplementary Section \ref{SM:second_order}) we can write the photoemission intensity of QP1 as:
\begin{equation}
    I^{\mathrm{H,L}}(E,\mathbf{k_{\parallel}}) \propto |\Delta_{f_D}^{\mathrm{H}} \Delta_{Di}^{\mathrm{L}}|^2 \ ,
\end{equation}
where H=S/P and $\mathrm{L} =\pm$ represent the polarization of the HE and LE photons respectively. $\Delta_{Di}^{\mathrm{L}}$ is the transition matrix element from initial state $i$ to the Dirac intermediate state $D$, and $\Delta_{fD}^{\mathrm{H}}$ is the matrix element from $D$ to a final state $f$ at in-plane momentum $\mathbf{k}_{\parallel}=(k_x,k_y)$.

Importantly, in the single-channel scenario, the contributions of the photoemission matrix element $\Delta_{fD}^{\mathrm{H}}(E, k_y)$ can be canceled out if we look at the normalized LE-CD:
\begin{align}
    I_\mathrm{CD_{norm}}^{\mathrm{H}}(E, k_y) & = \frac{I^{\mathrm{H,L=-}}-I^{\mathrm{H,L=+}}}{I^{\mathrm{H,L=-}}+I^{\mathrm{H,L=+}}} \\[6pt]
    &\propto \frac{|\Delta_{fD}^{\mathrm{H}}|^2\cdot( |\Delta_{Di}^{\mathrm{L=-}}|^2-|\Delta_{Di}^{\mathrm{L=+}}|^2)}{|\Delta_{fD}^{\mathrm{H}}|^2\cdot(| \Delta_{Di}^{\mathrm{L=-}}|^2+|\Delta_{Di}^{\mathrm{L=+}}|^2)} = \frac{|\Delta_{Di}^{\mathrm{L=-}}|^2-| \Delta_{Di}^{\mathrm{L=+}}|^2}{| \Delta_{Di}^{\mathrm{L=-}}|^2+| \Delta_{Di}^{\mathrm{L=+}}|^2}\nonumber.    
\end{align}

Therefore, if just one QP is present and there are no interference effects, $I_\mathrm{CD_{norm}}^{\mathrm{H}}(E, k_y)$ should be \textit{identical} for S- and P-polarized HE pulse, regardless of photoemission matrix elements, while any deviation from that indicates the contribution of a second channel.

The normalized-CD spectra for both S- and P-polarization of the HE pulse are shown in Fig.~\ref{fig:Interf_vs_matE}(a,b). Differences between the two spectra are clearly visible, demonstrating the contributions of interfering paths to the photoemission signal. To highlight the areas of the spectrum where the contribution of a second photoemission channel is significant we plot in panel (c) the k-symmetrized difference between the normalized CD-spectra: $|I_\mathrm{CD_{norm}}^{\mathrm{S}}(E, k_y)-I_\mathrm{CD_{norm}}^{\mathrm{P}}(E, k_y)|$. Here, the non-white areas represent the spectral regions that are influenced by QP interference, where the three most prominent features are marked by colored rectangles. 
This interpretation is further supported by two control measurements. First, the spectral signatures attributed to a second photoemission channel disappear at late pump-probe delays, when the coherent overlap between the two QPs is lost (Supplementary Section \ref{SM:time_dependence}). Second, the residual S/P photoemission matrix elements measured at these late times do not exhibit features that follow the Dirac-band dispersion (Supplementary Section \ref{SM:late_time_matrix_elements}), ruling out matrix-element effects as the origin of the observed band-following signatures.

These three regions all represent resonant transitions of QP1 as seen by the overlap of initial states (upshifted by LE photon energy, marked by dashed lines) and intermediate states (solid black lines). However, the excitation the electron undergoes in QP2 is quite different in each of the three regions. We consider each of them, from top to bottom.
In region 1 (blue square in Fig.~\ref{fig:Interf_vs_matE}(c)), the initial state lies only 5.94~eV below the vacuum level, so the HE photon ($\hbar\omega=5.96$~eV) in QP2 excites the electron above $E_\mathrm{vac}$ into the final state continuum (diagram \#1 in panel (d)). The IPS-based ON/OFF control is therefore lost, consistent with interference signatures appearing for both S- and P-polarized HE.
In region 2 (magenta square), the IPS overlaps energetically with the QP1 intermediate states, so QP2 is also resonant (diagram \#2 in panel (d)). This produces the strongest interference -- visible as destructive interference in the raw intensity (Fig.~\ref{fig:Interf_vs_matE}(e), magenta arrows) -- but is unsuitable for phase reconstruction because QP2 does not provide a constant phase reference, and the destructive interference obliterates much of the signal.

We now consider the lowest interference region (red rectangle in (c) and diagram \#3 in (d)). Here we have in QP1 exactly one initial state resonantly excited into one intermediate state -- which is moreover a Dirac topological surface state\cite{sobota_direct_2013}. QP2 here is far from resonance, therefore the phase accumulated along this path should have negligible energy or momentum dependence (see experimental validation in Supplementary Section~\ref{SM:QP2_sensitivity}). Here, QP2 can work well as the reference path for phase reconstruction of QP1, and the rest of this paper focuses on this region.

\subsection{Circular dichroism provides clear signature of interference}

\begin{figure}
\centering \includegraphics[width=\textwidth]{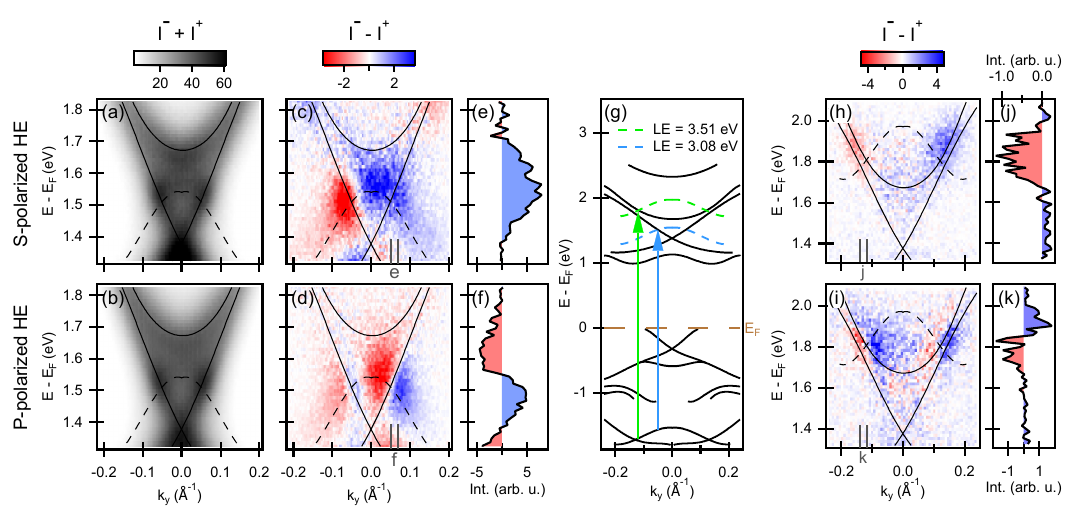}
\vspace{-15mm}
\caption{\textbf{Interference signature in CD spectrum}. (a-b) 2PPE intensity $I^- + I^+$ and (c-d) LE-CD spectra $I^- - I^+$, measured with LE = 3.08 eV, for S-polarized HE (a,c) and P-polarized HE (b,d). (e-f) EDCs of CD spectra integrated over the momentum range marked by gray lines in (c-d), respectively. (g) Schematic visualization of the optical transition from initial states to intermediate state of QP1 (black lines). The resonant transition for LE=3.08 eV (3.51 eV) is marked by a blue (green) arrow, with the corresponding upshifted initial state marked by a dashed line. (h-i) LE-CD spectra for LE = 3.51 eV, for S-polarized HE and P-polarized HE, respectively. (j-k) EDCs of CD spectra integrated over the momentum range marked by gray lines in (h-i), respectively. In all spectra, the intermediate states are marked by solid lines, and the upshifted initial states by dashed lines. 
}
\label{fig:CD_sum}
\end{figure}

Fig.~\ref{fig:CD_sum}(a) shows the one-QP trARPES spectrum (S-polarized HE) around the second Dirac cone, excited by a 3.08 eV LE pulse and 5.96 eV HE pulse. The two strong intensity peaks signify a resonant transition.  
The P-polarized HE case is shown in panel (b), and while we know from Fig.~\ref{fig:Interf_vs_matE}(c) that this data is affected by the interference between QP1 and QP2, the interference is not easily discernible in the intensity data.

We now proceed to explore more closely the CD of the signal in this region in both cases. In the one-path case (S-pol HE, Fig.~\ref{fig:CD_sum}(c)), the LE-CD is attributed to asymmetric excitation matrix elements due to the orbital angular momentum texture in the topological state\cite{soifer_band_2019}. The dichroic signal is maximal at the resonance, where most of the electrons originate from direct optical coupling rather than from secondary scattering processes, thereby conserving their momentum polarization. 
When examining the CD-spectrum of the two-path case (Fig.~\ref{fig:CD_sum}(d)), a drastic difference is observed: two additional sign changes appear, exactly on the centers of the bands (see also comparison between energy distribution curves (EDCs) in panels (e-f)). Such a sharp sign reversal across the band maximum is not expected from matrix-element effects, which typically vary smoothly with energy and momentum. This is also confirmed experimentally by late-time measurements, where only matrix-element effects remain and no band-following sign reversal is observed (Supplementary Section \ref{SM:late_time_matrix_elements}). Instead, we show that it arises naturally from interference between a resonant and a non-resonant pathway.
 
The sharp sign change across the band maximum is reminiscent of Fano-type signals, where an interference between a resonant transition and a continuum creates strong intensity modulation around the resonance energy\cite{fano_effects_1961, litvinenko_multi-photon_2021}. Similarly, in our case, there is one resonant path (QP1) and a second path with a flat-phase contribution. The intensity modulation in such cases stems from a $\pi$-phase-step, which is expected when crossing a resonance\cite{hazi_behavior_1979,dudovich_transform-limited_2001}, leading to a switch between constructive to destructive interference below and above the resonance. The phase-step is present in QP1 with or without the interference with QP2, however, it does not influence the measured intensity without the second path to interfere with, see simulation in Supplementary Section \ref{SM:CD_simulation}. 
This data is therefore a conclusive indication that our QP interferometer is sensitive to the phase accumulated in the resonant transition between the initial state and the unoccupied Dirac cone. Notably, this conclusion is obtained within a single HE polarization and therefore does not rely on comparisons between S and P matrix elements.

To confirm that the sign change is tied to the resonant transition in QP1, we tune the LE photon energy from 3.08 to 3.51 eV.
Fig.~\ref{fig:CD_sum}(g) shows the expected shift of the resonance both in energy and momentum, according to the dispersions of both initial and intermediate states. The one-path case (S-pol HE, Fig.~\ref{fig:CD_sum}(h,j)) displays a strong CD signal which has shifted as expected in comparison to panel (c). And indeed the two-path case (P-pol HE, panels (i,k)) displays a sign flip across the band maximum, at the new energy-momentum position of the resonance. This demonstrates that not only can we turn ON and OFF the interference by switching the HE pulse polarization from P to S, but we can also \textit{control} the energy-momentum position of the interference by tuning the LE resonance wavelength.

\subsection{Phase reconstruction}

Having established experimentally that the observed CD originates from coherent interference rather than matrix-element effects, we now use the ON/OFF reference arm to reconstruct the relative phase between the two QPs. For a given LE polarization, we measure the photoemission intensity with both S- and P- polarized HE pulse.
We denote by $\psi_1$ ($\psi_2$) the electronic wavefunction at the output of QP1 (QP2), and by $\Delta\phi$ the phase difference between them. 
The intensity for the two HE polarizations is therefore given by:
\[ I_S =  |A \cdot\psi_1|^2 \]
\[I_P = |\psi_1+\psi_2|^2 =  |\psi_1|^2+|\psi_2|^2+2 |\psi_1| |\psi_2|\cos(\Delta\phi),
\]
where $|A|$ is the matrix-element ratio between photoemission with S- and P-polarized HE pulses. 
Using these two equations, we can extract the phase difference $\Delta\phi$:
\begin{equation}
    \cos(\Delta\phi)=\frac{I_P-|\psi_1|^2-|\psi_2|^2}{2|\psi_1| |\psi_2|}=\frac{I_P-\left|\frac{1}{A}\right|^2I_S-|\psi_2|^2}{2\left|\frac{1}{A}\right|\sqrt {I_S} |\psi_2|},
\label{eq:cos}
\end{equation}
where $|\psi_2|^2\sim0.3$ is the intensity contribution of QP2 and $\left|1/A\right|^2\sim1.75$ (see Methods).  
In order to reconstruct $\cos(\Delta\phi)$, both the one-path and the two-path measurements are needed, highlighting the importance of the ability to turn ON/OFF the interference.

\begin{figure}
\centering \includegraphics[width=\textwidth]{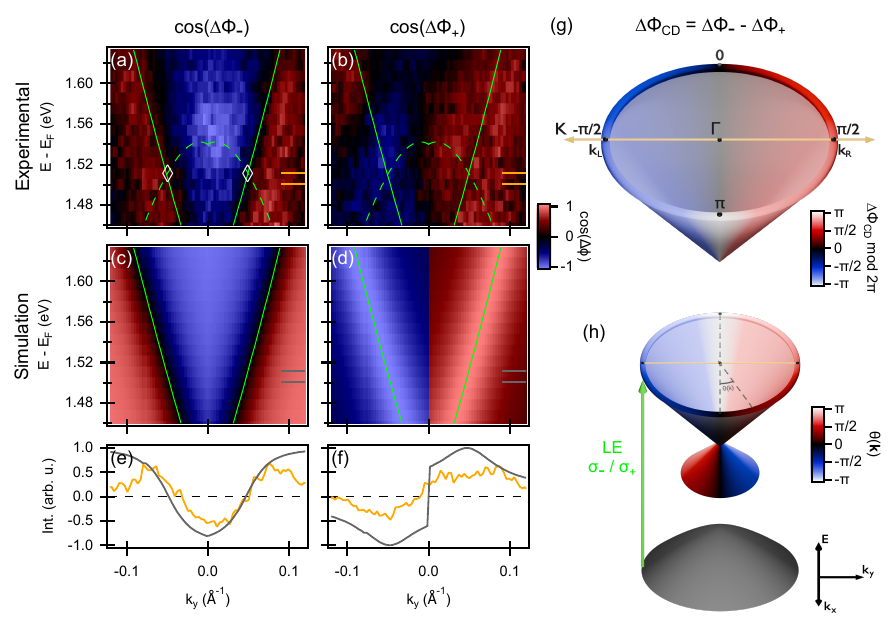}
\vspace{-13mm}
\caption{\textbf{Experimental phase reconstruction and relation to Dirac surface state.} (a-b) Experimental $\cos(\Delta\phi_-)$ and $\cos(\Delta\phi_+)$ respectively. (c-d) Simulated $\cos(\Delta\phi_-)$ and $\cos(\Delta\phi_+)$ from the phase in Eqs.~\eqref{eq:phase_rec_exp_min}-\eqref{eq:phase_rec_exp_plu}. Dispersion of the Dirac cone (intermediate state of QP1) is shown in solid lines in (a-d), dashed lines in (a-b) are the upshifted initial state. (e-f) Momentum Distribution Curves (MDCs) of experimental (gray) and simulated (orange) $\cos(\Delta\phi_\pm)$. Integration region is marked on the corresponding spectra (a-d). (g) Phase dichroism $\Delta \phi_{\mathrm{CD}}(\mathbf{k}_{\parallel})=\Delta \phi_-(\mathbf{k}_{\parallel})-\Delta \phi_+(\mathbf{k}_{\parallel})$ mod $2\pi$, along Dirac surface state. The yellow line denotes the measured $k_y$ path along $\Gamma-K$, where $k_{L/R}$ corresponds to the two experimental resonant $k_y$ points marked in white in (a). (h) The electronic transition amplitude phase. The LE excitation from the initial state (gray) to the Dirac cone is denoted by a green arrow. The phase of the Dirac surface state is defined with respect to the $x-y$ coordinate system so that $\tan(\theta(\textbf{k}_\parallel)) =k_y/k_x$.
}
\label{fig:4_phase}
\end{figure}

The interferometric measurement was performed independently for each of the two circular LE polarizations, and the reconstructed $\cos(\Delta\phi_-)$ and $\cos(\Delta\phi_+)$ (for $\sigma_-$ and $\sigma_+$ LE polarizations, respectively) are shown in Fig.~\ref{fig:4_phase}(a-b). We first focus on the interferometric data obtained with $\sigma_-$ polarized LE pulse, in panel (a). Notably, the $\cos(\Delta\phi_-)$ signal shows a clear sign change across the band center (marked in green), in agreement with a $\sim\pi$ phase shift across the resonance. Here we directly resolve this phase jump as it follows the band dispersion in energy and momentum, confirming our energy-momentum resolved interferometric scheme. An equivalent phase reconstruction is obtained with linearly polarized LE excitation (Supplementary Section \ref{SM:linear_LE}), demonstrating that the interferometric reconstruction does not rely on CD.

Next, we look at the data for $\cos(\Delta\phi_+)$ in Fig.~\ref{fig:4_phase}(b). Surprisingly, we see a different symmetry pattern: Whereas for $\sigma_-$ the pattern was symmetric, with the zero crossing along the Dirac band lines, here with $\sigma_+$ LE polarization, the $\cos(\Delta\phi_+)$ is anti-symmetric, with the maximum signal following the band line, while the zero crossing is at momentum $k_y\sim0$. We note however, that if we take a phase $\pi$-step which is shifted by $\pi/2$ relative to the previous case, then the cosine of this phase will be maximal on the resonance energy, and always positive. This can explain the fact that the maximum $\cos(\Delta\phi)$ is now on the band, but not, however, the change from symmetry to anti-symmetry. 

The markedly different symmetry of the $\sigma_-$ and $\sigma_+$ interferograms indicates that the phase difference between the two circular polarizations contains information beyond the resonant phase jump alone. We therefore define the phase dichroism,
\begin{equation}
\Delta\phi_{\rm CD}(E,\textbf{k}_\parallel)
\equiv
\Delta\phi_{-}(E,\textbf{k}_\parallel)-\Delta\phi_{+}(E,\textbf{k}_\parallel),
\end{equation}
which captures the polarization-dependent contribution to the measured phase. As shown in Supplementary Section~\ref{SM:second_order}, the contribution of the reference arm, QP2, cancels in this quantity, such that $\Delta\phi_{\rm CD}$ isolates the phase accumulated in the optical coupling to the Dirac intermediate state.

The wavefunction of a Dirac surface state is characterized by the spinor phase $\theta(\textbf{k}_\parallel)$, which encodes its helical spin- and orbital-angular-momentum texture\cite{PhysRevB.82.045122}, see Fig.~\ref{fig:4_phase}(g). Notably, the optical transition amplitudes inherit this phase structure, leading to
\begin{equation}
\Delta\phi_{\rm CD}(\textbf{k}_\parallel) -  \Delta\phi_{\rm CD}(-\textbf{k}_\parallel)=
\theta(\textbf{k}_\parallel) - \theta(-\textbf{k}_\parallel) = \pi \pmod{2\pi}.
\end{equation}
Thus, the phase dichroism provides experimental access to the helical phase texture of the Dirac surface state, as shown in Fig.~\ref{fig:4_phase}(h). For the reconstructed phase in Fig.~\ref{fig:4_phase}(a), we have $\Delta\phi_{\rm CD}=-\pi/2$ ($+\pi/2$) at the left (right) labeled $k_y=k_L$ ($k_y=k_R$) point. The measured phase dichroism is therefore consistent with the theoretical expectation above. We further plot the theoretical $\Delta\phi_{\rm CD}(\textbf{k}_\parallel)$ along the Dirac state at a constant energy in Fig.~\ref{fig:4_phase}(g), where the yellow line crossing the resonant points $k_L$ and $k_R$ is the experimentally measured momentum path. Any pair of $\pm \textbf{k}_\parallel$ points along the Dirac state has a $\pi$ difference in phase dichroism. This directly reflects the underlying phase $\theta(\textbf{k}_\parallel)$.

Guided by these observations, we now recover the underlying phase images that would lead to the $\cos(\Delta\phi)$ observed in the two connected data sets of Fig.~\ref{fig:4_phase}(a-b). For every value of $\cos(\Delta\phi)$, there are two possible values of phase, and we therefore seek a phase distribution that simultaneously reproduces the measured interferograms and the observed phase dichroism. With this requirement, the cosine in Fig.~\ref{fig:4_phase}(a-b) can be recreated -- including the symmetry change between the two polarizations -- for the following phase pattern:

\begin{equation}
    \label{eq:phase_rec_exp_min}
    \Delta\phi_-(E, k_y)=\arctan\!\left(\dfrac{E - \varepsilon(k_y)}{\gamma}\right) + \frac{\pi}{2}
\end{equation}
\begin{equation}
    \label{eq:phase_rec_exp_plu}
    \Delta\phi_+(E, k_y) =  
    \begin{cases}
        \Delta\phi_-(E, k_y) + \frac{\pi}{2}, & k_y<0\\[6pt]
        \Delta\phi_-(E, k_y) - \frac{\pi}{2}, & k_y\ge 0
    \end{cases}
\end{equation}

This phase is constructed such that for each momentum point, the energy-dependent phase is a $\pi$ phase step across the resonant energy at that momentum. The resonance energy per momentum is given by the Dirac band dispersion $\varepsilon(k_y)$, and the shape of the phase step is given by $\arctan(\frac{E-\varepsilon(k_y)}{\gamma})$ where $\gamma$ is the width of the resonance (the bandwidth)\cite{hazi_behavior_1979}.

The cosine images for the phases from Eqs.~\eqref{eq:phase_rec_exp_min}-\eqref{eq:phase_rec_exp_plu} are shown in Fig.~\ref{fig:4_phase}(c-d), respectively, with a direct comparison between data and reconstruction along the marked energy cut in panels (e-f). The reconstructed cosine of the phase presents an excellent agreement with the measured cosine data, capturing the main features: the symmetry to anti-symmetry switch between $\sigma_-$ to $\sigma_+$, the sign change across the band maximum in (a) and (c), and the sign change around $k_y=0$ in (b) and (d). 

\section{Discussion}

The phase reconstructed through our quantum-path interferometer reveals two distinct signatures: a resonance-associated phase evolution, and a momentum-odd phase dichroism linked to the helical Dirac cone.
The first signature is a $\pi$-phase shift associated with the resonant optical transition into the Dirac state. This phase evolution follows the Dirac-band dispersion and gives rise to the Fano-like dichroic lineshape observed in the interference signal. Fully resolving phase evolution across resonances remains an active topic in photonics and ultrafast science\cite{heeg_interferometric_2015,limonov_fano_2017,litvinenko_multi-photon_2021,zhang_resolving_2025}, and our results demonstrate that similar phase-sensitive phenomena can be measured directly in quantum materials with simultaneous energy and momentum resolution. Indeed, because the phase varies sharply in both energy and momentum, momentum-integrated measurements would strongly suppress the signal -- underscoring that energy- and momentum-resolved detection is essential for resolving such phase textures.

We also resolve a momentum-odd phase dichroism between the two circular polarizations of the LE excitation, with the relative phase shifted by $-\pi/2$ and $+\pi/2$ on opposite sides of the Dirac cone, consistent with the helical phase structure of the Dirac spinor. Recent works have demonstrated that CD-ARPES and related observables can uncover orbital and geometric information in quantum materials\cite{beaulieu_revealing_2020,schuler_polarization-modulated_2022,beaulieu_berry_2024,schusser_towards_2024}; our interferometric approach complements these efforts by providing direct access to the phase.
In our work, the strong spin-orbit coupling and resulting CD signal were instrumental in identifying and isolating the interference contribution, but are not a fundamental requirement of the interferometric scheme. The phase can be reconstructed independently for each circular polarization of the LE excitation, as well as for linearly-polarized excitation (Supplementary Section~\ref{SM:linear_LE}).

More broadly, the ability to measure phase textures in energy and momentum space opens new opportunities for studying quantum materials. While conventional photoemission probes spectral weight and population dynamics, interferometric photoemission provides access to phase information encoded in transition amplitudes. This capability may enable future studies of topological states, symmetry-breaking phases, and other phenomena where the phase structure of electronic states plays a central role. Furthermore, the approach could be extended to other forms of photoemission interference, including Floquet-Volkov-type processes\cite{mahmood_selective_2016, bao_floquet-volkov_2025, choi_observation_2025, merboldt_observation_2025, fragkos_floquet-bloch_2025}, provided suitable control schemes are developed.

The interferometric scheme is applicable to a broad range of material systems. Image-potential states are a generic feature of many surfaces\cite{echenique_induced_2002,borca_image_2024} and can provide a natural reference arm, while the excitation remains in the perturbative regime with photon energies in the visible-to-UV range. Importantly, the phase reconstruction is obtained directly from experimental observables and does not rely on a detailed model of the underlying electronic bands. Energy- and momentum-resolved electron interferometry, therefore, provides a direct route to accessing phase information in quantum materials. The practical implementation of the scheme does, however, require several conditions. A suitable intermediate state must act as a reference arm with a weak energy- and momentum-dependent phase, and must exhibit polarization-selective coupling that enables control over the second quantum-path. In addition, the photon energies must satisfy $\hbar\omega_{\mathrm{HE}}-\hbar\omega_{\mathrm{LE}}<\Phi$, ensuring that the non-resonant pathway remains below the vacuum level and does not couple directly to the final state continuum. These requirements are not specific to Bi$_2$Se$_3$ and are expected to be met in a wide range of materials, making the approach broadly applicable.

\section*{Methods}

\subsection{Experimental details}
TrARPES measurements were performed on \(\mathrm{Bi_2Se_3}\) single crystals along the $\Gamma-K$ momentum axis, with a DA30 hemispherical analyzer by ScientaOmicron. A sketch of the experimental geometry is shown in Fig.~\ref{fig:1}(b). 
The samples were cleaved cold \textit{in situ} and were kept at a temperature of 80K and pressure below $2\times10^{-10}$~mbar.
A bias voltage of -20 V between the sample and analyzer entrance was used to broaden the angular field of view\cite{pfau_low_2020,gauthier_expanding_2021, hengsberger_photoemission_2008}.
The translation of angle to momentum of the biased spectra was performed according to Gauthier \textit{et al.}\cite{gauthier_expanding_2021}.
In all of the main text measurements, the HE pulse energy was 5.96 - 6.00 eV and linearly polarized in either P- or S-polarization (in the plane of incidence or in the plane of the sample surface, respectively, see Fig.~\ref{fig:1}(b)).
The LE pulse was circularly polarized ($\sigma_\pm$) and at 3.08 eV photon energy. Only the measurements of Fig.~\ref{fig:CD_sum}(h-k) were performed with LE=3.51 eV.

The requirement that QP2 is non-resonant with the intermediate state means that the energy of the initial state plus one HE photon should still be within the energy range of the material's bound states ($E_\mathrm{inter}=E_\mathrm{init}+\hbar\omega_\mathrm{HE}<E_\mathrm{vac}$), since in the final state continuum above $E_{\mathrm{vac}}$ all transitions are resonant. Combined with the standard condition that the intermediate states should be unoccupied, hence $E_\mathrm{inter}>E_\mathrm{F}$ for both QP1 and QP2, this requirement sets a limit on the energy difference between LE and HE pulses:  $\hbar\omega_{\mathrm{HE}}-\hbar\omega_\mathrm{LE}<E_\mathrm{vac}-E_\mathrm{F}$, where $\Phi=E_\mathrm{vac}-E_\mathrm{F}$ is the work-function of the material. This requirement is central for the implementation of the interferometric scheme, as it ensures that QP2 remains non-resonant and can serve as a phase reference.
The work-function of the samples used in our experiments is $\Phi=5.44$ eV, and therefore the energy difference of $\sim3$ eV between the pulses fits the requirement.

The dataset used for phase reconstructions (data displayed in: Fig.~\ref{fig:Interf_vs_matE}, Fig.~\ref{fig:CD_sum}(a-f), Fig.~\ref{fig:4_phase}(a-b) and Fig.~\ref{fig:bands}(a)) was measured with LE pulse duration of 130 fs, with an overall cross-correlation time resolution of 147 fs. 
The pump-probe delay of $\Delta t=+70$ fs was chosen according to the time delay where the intensity at the LE resonance ($E-E_\mathrm{F}=1.55$~eV) is maximal. Time-dependent measurements show that the interference contribution remains significant throughout the pulse-overlap region and that the reconstructed phase is unchanged, within experimental resolution, between zero delay and the signal maximum (Supplementary Section \ref{SM:time_dependence}).
Similarly, the CD data for LE=3.51~eV (Fig.~\ref{fig:CD_sum}) is measured at the delay where the signal at $E-E_\mathrm{F}=1.8$~eV (resonance position) is maximal (pulse duration of 70 fs and $\Delta t=20$~fs). 

The spectrum of Fig. \ref{fig:1}(d) is at $\Delta t=-25$ fs, chosen so that both the IPS and intermediate states of QP1 are clearly visible. 

\begin{figure*}
\centering \includegraphics[width=0.5\textwidth]{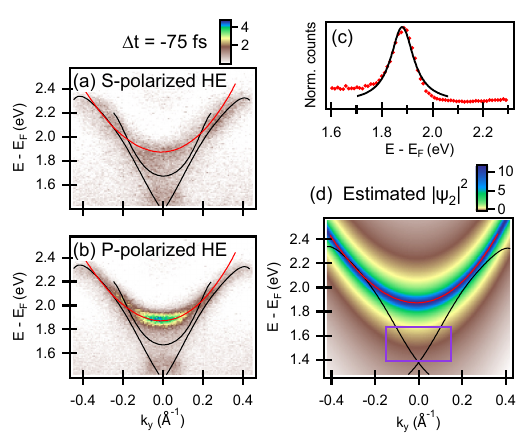}
\vspace{-6mm}
\caption{\textbf{Image Potential State.} Final states spectrum of \(I^- + I^+\), at $\Delta t = -75$ fs, for (a) S-polarized (b) P-polarized HE. The QP1 intermediate states are marked by a solid black line, the IPS (QP2 intermediate state) is marked by a solid red line. (c) EDC fit (solid black line) of the center of the IPS band, red dots are the data. (d) The estimated intensity from QP2, $|\psi_2|^2$, in logarithmic scale. The purple rectangle marks the region used for $\cos(\Delta\phi)$ reconstruction (Fig.~\ref{fig:4_phase}).}
\label{fig:IPS}
\end{figure*}

\begin{figure}
\centering \includegraphics[width=0.9\textwidth]{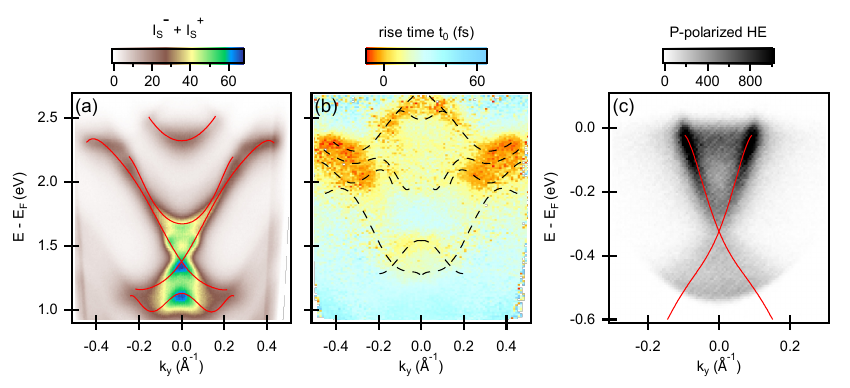}
\vspace{-6mm}
\caption{\textbf{Band dispersions.} (a) $I^- + I^+$ measured with S-polarized HE (the full range spectrum from Fig.~\ref{fig:CD_sum}(a)). The QP1 intermediate states lines used throughout the manuscript are in red. (b) rise time in energy and momentum space of $I^- + I^+$ done with S-polarized HE. The initial states (dashed black lines) are drawn according to early rise time analysis\cite{soifer_band_2019}. (c) Spectrum of the first Dirac cone measured by P-polarized HE via ARPES. The solid red lines are the continuation of the dashed line in (b).}
\label{fig:bands}
\end{figure}

The matrix-element ratio between photoemission with S- and P-polarized HE pulses ($|A|$ in Eq. \ref{eq:cos}) is estimated by taking the ratio $\frac{I^-_S + I^+_S}{I^-_P + I^+_P}$ for a band showing no resonance signal (at $E-E_F\sim1$~eV), giving a value of $\frac{1}{|A|^2}\sim1.75$. Late-time measurements further confirm that this ratio is nearly independent of energy and momentum within the phase-reconstruction region (Supplementary Section \ref{SM:late_time_matrix_elements}). The intensities shown for S-polarized HE pulse in Fig.~\ref{fig:Interf_vs_matE}(e), Fig.~\ref{fig:CD_sum}(a,c), Fig.~\ref{fig:IPS}(a)
and Fig.~\ref{fig:bands}(a) are already divided by $|A|^2$. 

\subsection{Image Potential State}

The strong polarization dependence of the coupling to the IPS can be seen in Fig.~\ref{fig:IPS}(a-b). The spectra are shown at a negative time delay, where due to the reversed pump-probe roles the IPS is clearly visible and the rest of the intermediate states are suppressed. The IPS intensity at P polarization (b) is roughly 5 times stronger than at S polarization (a). 

Fig.~\ref{fig:IPS}(d) shows the estimated $|\psi_2(E, k_y)|^2$, which is a Lorentzian shape in energy around the maximum located at IPS position (red solid line in (a) and (b)), where the Lorentzian width $\gamma=55$~meV is obtained from fitting the IPS lineshape from (b) (see fit in panel (c)).  The overall amplitude is estimated as 11.2, chosen as the minimal amplitude that can agree with the interference signature we observe. The magnitude within the energy-momentum region where we reconstruct the phase (purple square) is 0.2-0.4 (intensity is relative to the counts reported in Fig.~2-4).  We note that changes in $|\psi_2(E, k_y)|^2$, such as taking a constant $|\psi_2(E, k_y)|^2=0.25$ or increasing the overall amplitude does not change the resulting phase reconstruction significantly. We further verified the robustness of the reconstruction experimentally by repeating the analysis with a shifted HE photon energy, thereby changing the IPS detuning. The reconstructed phase remained unchanged within experimental resolution (Supplementary Section \ref{SM:QP2_sensitivity}).

\subsection{Band dispersions}
Band dispersions for initial and intermediate states used throughout the manuscript are extracted from our data (as detailed below), in agreement with previous experiments and DFT calculations\cite{soifer_band_2019, sobota_direct_2013,aguilera_many-body_2019, sidilkover_reexamining_2025}. 
Intermediate states in the energy range of 1-2.5 eV were extracted from the measured trARPES spectrum at S polarization (QP1 only), see Fig.~\ref{fig:bands}(a). The IPS dispersion was extracted from trARPES spectrum at P polarization at early times showing the IPS (Fig.~\ref{fig:IPS}(b)). Initial states dispersions are extracted by tracking the early-rise-time regions in our time-resolved data (Fig.~\ref{fig:bands}(b)), according to the time-mapping procedure outlined in\cite{soifer_band_2019}, with guidance also from DFT calculations and synchrotron data\cite{sidilkover_reexamining_2025}.
The first Dirac cone is directly measured by our experiment setup with 6 eV photoemission, as seen in Fig.~\ref{fig:bands}(c). The red (solid) Dirac band lines are the continuation of those (dashed) in Fig.~\ref{fig:bands}(b). The work function $\Phi$ is calculated according to the low-energy cutoff seen in this spectrum. 

\section*{Acknowledgments}
We thank A. Ron, H. Suchowski, R. Ilan and D. Azoury for valuable discussions. I. S. acknowledges funding from the National Quantum Science and Technology Program of the Israeli Planning and Budgeting Committee. H. S. acknowledges the support of the Zuckerman STEM Leadership Program, the Young Faculty Award from the National Quantum Science and Technology program of the Israeli Planning and Budgeting Committee, the ERC PhotoTopoCurrent 101078232 and the Israel Science Foundation (grant no. 2117/20). Y. Y. and M. A. S. acknowledge funding by the Deutsche Forschungsgemeinschaft (DFG, German Research Foundation) through project no.~531215165 (Research Unit OPTIMAL).

\bibliography{Bibliography}

\clearpage

\begin{supp}

\begin{center}
    {\large Supplementary Material: Resolving the phase of a Dirac topological state via interferometric photoemission}
\end{center}

\setcounter{page}{1} 
\renewcommand{\thepage}{S\arabic{page}} 

\setcounter{figure}{0}
\renewcommand{\thefigure}{S\arabic{figure}}
\renewcommand{\theHfigure}{S\arabic{figure}}

\setcounter{table}{0} 
\renewcommand{\thetable}{S\arabic{table}} 
\renewcommand{\theHtable}{S\arabic{table}}

\setcounter{equation}{0} 
\renewcommand{\theequation}{S\arabic{equation}}
\renewcommand{\theHequation}{S\arabic{equation}}

\setcounter{section}{0}
\renewcommand{\thesection}{\Roman{section}}

\makeatletter
\setcounter{secnumdepth}{2}
\def\@seccntformat#1{\csname the#1\endcsname.\quad}
\makeatother

\def\@seccntformat#1{\csname the#1\endcsname.\quad}

\noindent\textbf{This PDF file includes:} \vspace{2mm}\\
\noindent Supplementary Text \\
\noindent Figures S1 to S6 \\
\noindent Equations S1 to S17 \\
\noindent References [57--61]

\newpage

\section{Second-order perturbation analysis of the interferometric photoemission}\label{SM:second_order}

The quantum-path (QP) interference in two-photon photoemission (2PPE) can be analyzed formally with second-order perturbation theory. We consider a time-dependent perturbation which is adiabatically switched on, combining two optical fields: a low-energy (LE) pulse with photon energy $\omega_L$, and a high-energy (HE) pulse with photon energy $\omega_H$, as illustrated in Fig.~1(a) in the main text. Throughout this section, we use atomic units and set $\hbar=1$. The light-matter interaction is described by the perturbation:
\begin{align}
    \label{eq:second_perturb}
    \hat{V}(t) = e^{\eta t}\left( \hat{\Delta}_L e^{-i\omega_L t} + \hat{\Delta}_L^\dagger e^{+i\omega_L t} + \hat{\Delta}_H e^{-i\omega_H t} + \hat{\Delta}_H^\dagger e^{+i\omega_H t}\right),
\end{align}
where $\eta>0$ is an infinitesimal adiabatic switching parameter that phenomenologically accounts for the finite pulse envelope. The operators $\hat{\Delta}_{H,L}$ denote the dipole coupling associated with the LE and HE fields, including their polarization dependence. The second-order transition coefficient from an initial state $\ket{i}$ into a final state $\ket{f}$ is,
\begin{align}
    \label{eq:transition_amplitude}
    c_{f}^{(2)}(t=0) = (-i)^2 \int_{-\infty}^{0} dt_1 \int_{-\infty}^{t_1} dt_2 \sum_n \langle f | \hat{V}_I(t_1) | n \rangle \langle n | \hat{V}_I(t_2) | i \rangle ,
\end{align}
where $\hat{V}_I$ refers to the perturbation in Eq.~\eqref{eq:second_perturb} in the interaction picture. Eq.~\eqref{eq:transition_amplitude} sums over all possible intermediate states $\ket{n}$. The dominant contributions are those bands that are closest to being resonant with the excitation energy. In our experiment, these consist of two contributions: QP1, in which the unoccupied Dirac state $\ket{D}$ is resonantly excited by $\omega_L$, and QP2, in which the IPS state -- though detuned -- is the closest band to the $\omega_H$ transition. 
Retaining only these two contributions, the transition rate from initial state with energy $E_i$ to final state with energy $E$ is simplified as a Fermi's golden rule with second-order matrix elements, 
\begin{align}
     \Gamma_{i \rightarrow f}(E,\mathbf{k}_\parallel) \propto |A_{fi}(E,\mathbf{k}_\parallel)|^2 \delta(\omega_{fi} - \omega_L - \omega_H) ,
\end{align}
where $\omega_{fi} \equiv E-E_i$ is the energy difference between initial and final states and $\mathbf{k}_\parallel=(k_x,k_y)$ is the in-plane crystal momentum defined along the crystal surface. We decompose the transition amplitude into two QPs as:
\begin{align}
    \label{eq:second_me}
    A_{fi}(E,\mathbf{k}_\parallel)  = \psi_{\mathrm{1}}^{\mathrm{H,L}}(E,\mathbf{k}_\parallel) + \psi^{\mathrm{H,L}}_{\mathrm{2}}(E,\mathbf{k}_\parallel)  ,
\end{align}
where $\psi_1^{\mathrm{H,L}}$ and $\psi_2^{\mathrm{H,L}}$ correspond to the outputs of QP1 and QP2, respectively. This is the same notation as used to describe the interferometer in the main text. $\mathrm{H}$ and $\mathrm{L}$ label the polarization for HE and LE photons. With second-order perturbation theory, they are evaluated as: 
\begin{align}
    \label{eq:QP12}
    \psi_1^{\mathrm{H,L}}(E,\mathbf{k}_\parallel)= \frac{\Delta_{fD}^{\mathrm{H}}\Delta_{Di}^{\mathrm{L}}}
    {\eta+i(\omega_{Di}-\omega_L)} \ , \qquad
    \psi_2^{\mathrm{H,L}}(E,\mathbf{k}_\parallel)=
    \frac{\Delta_{fP}^{\mathrm{L}}\Delta_{Pi}^{\mathrm{H}}}
    {\eta+i(\omega_{Pi}-\omega_H)}\ ,
\end{align}
where $\Delta_{ab}^{\mathrm{H/L}} \equiv \langle a (\mathbf{k}_\parallel) | \hat{\Delta}_{\mathrm{H}/\mathrm{L}} |b (\mathbf{k}_\parallel) \rangle$ are the polarization-dependent dipole matrix elements, and we omit the momentum label for simplicity. Eq.~\eqref{eq:second_me} captures the essential physics of the interferometer: the measured photoemission signal is the coherent sum of a resonant pathway (QP1) and a weakly dispersive reference pathway (QP2). The resonant denominator in QP1 gives rise to the characteristic phase evolution across the Dirac resonance, while the off-resonant QP2 acts as an approximately constant phase reference. In general, the two QPs can end in different final states. Rather than assuming identical final states, we show below that the experimentally reconstructed phase is largely insensitive to their contribution.

The physical role of the two pathways is transparent: $\psi_1$ is the resonant pathway mediated by the Dirac intermediate state, whose resonant denominator generates the characteristic arctangent phase evolution across the resonance, while $\psi_2$ is the IPS-mediated reference pathway. In the phase-reconstruction region discussed in the main text, this pathway is sufficiently detuned that its phase varies only weakly with energy and momentum.

In the following, we distinguish between three different phases: (i) the phase difference between the two QPs measured by the interferometer, (ii) the phase of the optical transition amplitudes into the Dirac state, and (iii) the intrinsic phase of the Dirac spinor. The goal of this section is to show how these quantities are related.

\subsection{Phase reconstruction}
The photoemission intensity measured with P-polarized HE excitation can therefore be written as: 

\begin{align}
\label{eq:SM_I_P}
I^{\mathrm{H}=\mathrm{P},\mathrm{L}}(E,\mathbf{k}_{\parallel}) \propto  & |\psi_{\mathrm{1}}^{\mathrm{H}=\mathrm{P},\mathrm{L}}(E,\mathbf{k}_{\parallel})|^2 + |\psi_{\mathrm{2}}^{\mathrm{H}=\mathrm{P},\mathrm{L}}(E,\mathbf{k}_{\parallel})|^2 \\ & + 2 |\psi_{\mathrm{1}}^{\mathrm{H}=\mathrm{P},\mathrm{L}}(E,\mathbf{k}_{\parallel})||\psi_{\mathrm{2}}^{\mathrm{H}=\mathrm{P},\mathrm{L}}(E,\mathbf{k}_{\parallel})| \cos[\Delta \phi_{\mathrm{L}}(E,\mathbf{k}_{\parallel})] \nonumber  .
\end{align}
Here $\Delta\phi_\mathrm{L}=\arg \left[\psi_1^\mathrm{H=P,L}\psi_2^{\mathrm{H=P,L}}\right]=\arg[\psi_1^\mathrm{H=P,L}]-\arg[\psi_2^\mathrm{H=P,L}]$  is the phase difference between the two QPs, and $\cos[\Delta\phi_\mathrm{L}(E,k_y)]$ is the quantity we reconstruct in our experiment.

In order to obtain $|\psi_1^{\mathrm{H}=\mathrm{P},\mathrm{L}}(E,\mathbf{k}_{\parallel})|^2$, we measure the intensity with S-polarized HE where only QP1 is active and we have,
\begin{align}\label{eq:SM_I_S}
    I^{\mathrm{H}=\mathrm{S},\mathrm{L}}(E,\mathbf{k}_{\parallel}) \propto |\psi_{\mathrm{1}}^{\mathrm{H}=\mathrm{S},\mathrm{L}}(E,\mathbf{k}_{\parallel})|^2. 
\end{align}

Next, we assume that, 
\begin{align}
\label{eq:SM_A_ratio}
    |A|^2|\psi_{\mathrm{1}}^{\mathrm{H}=\mathrm{P},\mathrm{L}}(E,\mathbf{k}_{\parallel})|^2 = |\psi_{\mathrm{1}}^{\mathrm{H}=\mathrm{S},\mathrm{L}}(E,\mathbf{k}_{\parallel})|^2 \ ,
\end{align}
where $A$ is a proportionality factor between the matrix elements (measured independently; see Methods) with negligible energy-momentum dependence in the reconstruction region (see experimental validation in section \ref{SM:late_time_matrix_elements}). 
Using Eqs.~\eqref{eq:SM_I_P}-\eqref{eq:SM_A_ratio} we reconstruct $\cos(\Delta\phi_\mathrm{L})$ from the measured intensities (Eq.~\eqref{eq:cos} in main text).

The experimentally reconstructed quantity is therefore determined by the phases of the two second-order transition amplitudes. Using Eqs.~\eqref{eq:second_me} and \eqref{eq:QP12}, the phase difference between the two QPs near the resonance energy ($E=E_r$) can be approximated as:
 
\begin{align}
    \label{eq:phase_pm}
    \Delta \phi_{\mathrm{L}=\pm}(E=E_r,\mathbf{k}_{\parallel}) \approx \arg[\Delta_{fP}^{\mathrm{L}=\pm} \Delta_{Pi}^{\mathrm{H=P}}]  - \frac{\pi}{2}  - \arg[\Delta_{fD}^{\mathrm{H=P}} \Delta_{Di}^{\mathrm{L=\pm}}]  - \arg\left[\frac{1}{\eta + i(\omega_{Di}-\omega_L)}\right],
\end{align}
where $\pi/2$ comes from the large detuning condition $\omega_{Pi} - \omega_H \gg \eta$ , which is satisfied in the phase-reconstruction region (interference region \#3 in Fig.~2(c) of the main text). The circularly polarized dipole operators can be decomposed as $\Delta^{\pm}=\Delta^{\mathrm{S}} \pm i\Delta^{\mathrm{P}}$. 
Using the experimentally observed selectivity of QP2 (due to IPS selection rules), $|\Delta_{fP}^{\mathrm{S}}| \ll |\Delta_{fP}^{\mathrm{P}}|$, the reconstructed phase simplifies to:
\begin{align}
    \label{eq:phase_pm2}
    \Delta \phi_{\mathrm{L}=\pm}(E=E_r,\mathbf{k}_{\parallel}) \approx \pm\frac{\pi}{2} + \arg[\Delta_{fP}^{\mathrm{L=P}} \Delta_{Pi}^{\mathrm{H=P}}] - \frac{\pi}{2} - \arg[\Delta_{fD}^{\mathrm{H=P}}\Delta_{Di}^{\mathrm{L=\pm}} ] - \arg\left[\frac{1}{\eta + i(\omega_{Di}-\omega_L)}\right].
\end{align}
Within these approximations, Eq.~\eqref{eq:phase_pm2} shows that the phase difference between the two circular polarizations is governed primarily by the matrix elements coupling the initial state to the Dirac intermediate state. 

We now define the \textit{phase dichroism} as:
\begin{align}
    \label{eq:phase_difference}
    \Delta \phi_\mathrm{CD}(\mathbf{k}_{\parallel}) \equiv \Delta \phi_{\mathrm{L}=-}(\mathbf{k}_{\parallel}) - \Delta \phi_{\mathrm{L}=+}(\mathbf{k}_{\parallel}) = -\pi + \arg[\Delta_{Di}^{+}(\mathbf{k}_{\parallel})]-\arg[\Delta_{Di}^{-}(\mathbf{k}_{\parallel})] \ .
\end{align}
For simplicity, the energy $E=E_r$ is dropped. Importantly, although the final photoelectron states associated with QP1 and QP2 may differ\cite{strocov_high-energy_2023}, the phase dichroism eliminates contributions from the final states and from the IPS-mediated reference pathway. The experimentally reconstructed phase difference therefore directly probes the phase structure associated with optical coupling into the Dirac state.

\subsection{Relation between the reconstructed phase and the phase of the Dirac state}
In this subsection, we relate the experimentally reconstructed phase to the intrinsic phase of the Dirac surface state wavefunction. The derivation proceeds in two steps. First, we express the optical transition matrix element in terms of the Dirac spinor. We then show that the experimentally reconstructed phase dichroism isolates the spinor phase.

The topological surface state of a three-dimensional topological insulator may be described by the effective Dirac Hamiltonian\cite{PhysRevB.82.045122, zhang2009topological}  
\begin{align}
    \label{eq:minimal_2orb_ham}
    H(\mathbf{k}_{\parallel}) \sim A_0 (\sigma_x k_y - \sigma_y k_x),
\end{align}
where $\sigma_x,\sigma_y$ are Pauli matrices and $A_0$ is a material-dependent constant. The Hamiltonian captures the helical spin texture and spin-momentum locking characteristic of the Dirac surface state. The Hamiltonian can be used to model the unoccupied Dirac state due to its similar spin texture\cite{sobota_direct_2013, niesner2012unoccupied}. The basis states $\ket{+}$ and $\ket{-}$ of the Hamiltonian in Eq.~\eqref{eq:minimal_2orb_ham} have well-defined $J=\pm \frac{1}{2}$ quantum numbers, respectively, in the presence of strong spin-orbit coupling. In particular, $\ket{+}$ ($\ket{-}$) is  a superposition of atomic orbitals $p_z$ and $p_+ = (p_x+ip_y)/\sqrt{2}$ ($p_z$ and $p_- = (p_x-ip_y)/\sqrt{2}$)\cite{PhysRevB.82.045122}. The upper Dirac eigenstate can then be solved as:  
\begin{align}
    \label{eq:dirac_wavefunc}
    \ket{D(\mathbf{k}_{\parallel})} = \frac{1}{\sqrt{2}}\left(\ket{+(\mathbf{k}_{\parallel})}-ie^{i\theta(\mathbf{k}_{\parallel})}\ket{-(\mathbf{k}_{\parallel})}\right) ,
\end{align}
where the $\mathbf{k}_{\parallel}$ dependent orbitals are obtained by Fourier transformation of the atomic orbitals. $\theta(\mathbf{k}_{\parallel})=\tan^{-1}(k_y/k_x)$ is the momentum-dependent spinor phase, which gives rise to the helical Dirac texture. The Dirac state in Eq.~\eqref{eq:dirac_wavefunc} exhibits nontrivial topology, characterized by a Berry phase of $\pi$~\cite{xiao2010berry} acquired upon encircling the Dirac point. The optical matrix element coupling the initial state $\ket{\psi_i(\mathbf{k}_\parallel)}$ to the Dirac state $\ket{D(\mathbf{k}_{\parallel})}$ is evaluated as,

\begin{align}
    \Delta_{Di}^{\mathrm{L}}(\mathbf{k}_{\parallel}) = \frac{1}{\sqrt{2}} (\langle +(\mathbf{k}_{\parallel})|\hat{\Delta}^{\mathrm{L}} | \psi_{i}(\mathbf{k}_{\parallel})\rangle -ie^{i\theta(\mathbf{k}_{\parallel})} \langle -(\mathbf{k}_{\parallel})|\hat{\Delta}^{\mathrm{L}} | \psi_{i}(\mathbf{k}_{\parallel})\rangle)  .
\end{align}
For circular polarization $\mathrm{L} =+$ ($\mathrm{L} =-$), optical selection rule allows the transition from initial state $s$ orbital to $p_+$ ($p_-$) in the Dirac state. Therefore we can evaluate the argument differences of the matrix elements between $\mathrm{L} =\pm$ as,
\begin{align}
    \label{eq:arg_difference}
    \arg[\Delta_{Di}^{+}(\mathbf{k}_{\parallel})] - \arg[\Delta_{Di}^{-}(\mathbf{k}_{\parallel})] = \arg[\langle +(\mathbf{k}_{\parallel})|\hat{\Delta}^{+} | \psi_{i}(\mathbf{k}_{\parallel})\rangle] - \theta(\mathbf{k}_\parallel) - \arg[i\langle -(\mathbf{k}_{\parallel})|\hat{\Delta}^{-} | \psi_{i}(\mathbf{k}_{\parallel})\rangle]  .
\end{align}
Interestingly, Eqs.~\eqref{eq:arg_difference} and ~\eqref{eq:phase_difference} demonstrate that the experimentally reconstructed phase dichroism is directly sensitive to the intrinsic phase $\theta(\mathbf{k}_{\parallel})$ of the Dirac spinor. 

In general, Eq.~\eqref{eq:arg_difference} would need to be modified for different experimental geometries, but $\Delta_{Di}(\mathbf{k}_{\parallel})$ is still determined by the phase $\theta(\mathbf{k}_{\parallel})$. While the phase of the Dirac state $\theta(\mathbf{k}_\parallel)$ is gauge and coordinate dependent, phase differences between different momenta are uniquely defined. To connect to the experimentally reconstructed phase, we compare Eq.~\eqref{eq:arg_difference} at opposite momenta $\pm \mathbf{k}_{\parallel}$. Considering inversion symmetry of the dipole matrix elements, 
\begin{align}
    \label{eq:arg_minusk}
    \arg&[\Delta_{Di}^{+}(-\mathbf{k}_\parallel)] - \arg[\Delta_{Di}^{-}(-\mathbf{k}_\parallel)] \nonumber \\&= \arg[\langle +(\mathbf{-k}_\parallel)|\hat{\Delta}^{+} | \psi_{i}(\mathbf{k}_\parallel)\rangle] - \theta(-\mathbf{k}_\parallel) - \arg[i\langle -(-\mathbf{k}_\parallel)|\hat{\Delta}^{-} | \psi_{i}(-\mathbf{k}_\parallel)\rangle] \nonumber , \\
    &= \arg[(-1)\langle +(\mathbf{k}_\parallel)|\hat{\Delta}^{+} | \psi_{i}(\mathbf{k}_\parallel)\rangle] - \theta(\mathbf{k}_\parallel) - \pi - \arg[(-i)\langle -(\mathbf{k}_\parallel)|\hat{\Delta}^{-} | \psi_{i}(\mathbf{k}_\parallel)\rangle]  \ .
\end{align}

This symmetry allows us to directly link the phase dichroism to the spinor phase $\theta$:
\begin{align}
    \Delta \phi_\mathrm{CD}(\mathbf{k}_\parallel) - \Delta \phi_\mathrm{CD}(-\mathbf{k}_\parallel) = \theta (\mathbf{k}_\parallel) - \theta (-\mathbf{k}_\parallel)=\pi \pmod{2\pi} \ ,
\end{align}
which is consistent with the experimental phase reconstruction in Fig.~4 of the main text. 

To summarize, the experimentally reconstructed phase should therefore be viewed as a phase of the optical transition amplitude into the Dirac state, rather than a direct measurement of the many-body Bloch wavefunction itself. Nevertheless, because the transition matrix elements inherit the spinor structure and wavefunction of the Dirac state, the measured phase retains direct sensitivity to the topological helicity encoded in the surface-state wavefunction. 

\section{Temporal dependence of population and interference signals}\label{SM:time_dependence}

To distinguish between incoherent population dynamics and coherent QP interference, we analyze the temporal evolution of both the total photoemission intensity and the CD signal. The measurements presented in this section were performed with a different dataset than the one used in the main text, with a shorter pulse duration and a cross-correlation time resolution of approximately 70 fs. This improved temporal resolution allows for a clearer separation between population dynamics and coherence effects.

\begin{figure}
\centering
\includegraphics[width=0.9\textwidth]{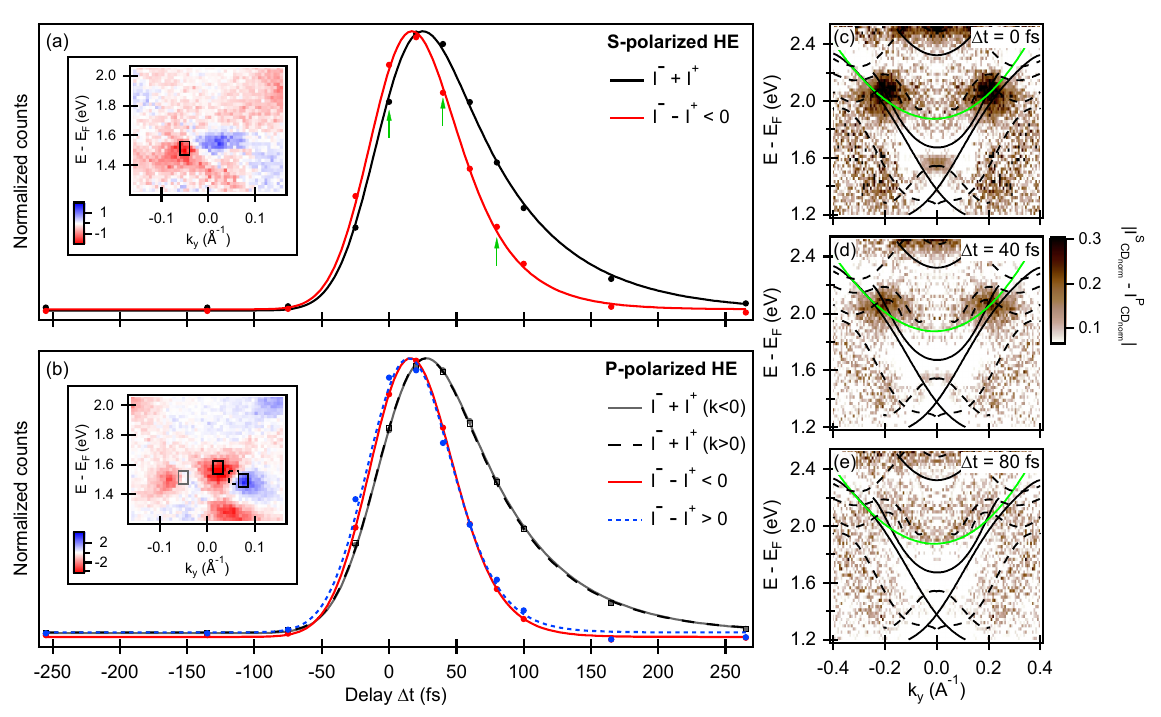}
\vspace{-5mm}
\caption{\textbf{Temporal behavior of $I^-+I^+$ vs $I^--I^+$}. In (a-b) both the data (dots) and the fitted time shape (line) are normalized to the data's maximum value. (a) S-polarized HE: $I^-+I^+$ (black) and $I^--I^+$ (red) taken from the same range shown in the inset CD spectrum. A fit to a gaussian convoluted with exponential decay resulted in a FWHM of $61\pm3$~fs ($61\pm3$~fs) and a $\tau$ of $60\pm4$~fs ($32\pm3$~fs) for $I^-+I^+$ ($I^--I^+$). (b) P-polarized HE: gray is $I^-+I^+$ from the same region as in (a), dashed black is for the green region in the inset CD spectrum. The red (blue) $I^--I^+$ corresponds to the black rectangle around the red (blue) signal in the CD inset. The temporal fit resulted in a FWHM of $65\pm1$~fs ($65\pm2$~fs) and a $\tau$ of $56\pm2$~fs ($54\pm2$~fs) for $I^-+I^+(k<0)$ ($I^-+I^+(k>0)$). While $I^--I^+<0$ ($I^--I^+>0$) resulted in a FWHM of $65\pm6$~fs ($65\pm1$~fs) and a $\tau$ of $18\pm6$~fs ($19\pm1$~fs). (c-e) $|I_\mathrm{CD_{norm}}^{\mathrm{S}}-I_\mathrm{CD_{norm}}^{\mathrm{P}}|$ at $\Delta t=0$, $+40$~and $+80$~fs respectively (times marked by green arrows in (a)).}
\label{fig:SM_CD_TDC}
\end{figure}

Figure~\ref{fig:SM_CD_TDC} presents the time dependence of the photoemission intensity $I^{-} + I^{+}$ (a) and the dichroic signal $I^{-} - I^{+}$ (b) for both S- and P-polarized HE excitation. We fit the dynamics to a sharp rise followed by exponential decay (representing the intrinsic dynamics of photoexcited electrons) convoluted with a gaussian representing the temporal resolution\cite{soifer_band_2019}. The total population, in either S or P, exhibits a characteristic decay with a lifetime of $\tau \sim 60$~fs, reflecting the relaxation dynamics of the excited electronic states.
In contrast, the dichroic signal shows a significantly faster decay. For S-polarized HE excitation (single-path case), the CD signal decays with a shorter lifetime of $\sim 30$~fs, providing a measure of the coherence time associated with the optically excited states. For P-polarized HE excitation (two-path case), the interference-related CD features exhibit an even faster decay, with a characteristic timescale of $\sim 20$~fs, and closely follow the temporal overlap of the two pulses.

This behavior indicates that the interference signal is governed by the coherent overlap of the excitation pathways and disappears once temporal overlap is lost, even though electronic population in the excited states persists for significantly longer times.

To further quantify this, we examine the normalized difference between the CD signals measured with S- and P-polarized HE excitation, which serves as a proxy for the contribution of the second QP. This quantity is maximal near zero delay (c), where the pulses overlap, and remains finite throughout the temporal overlap region, including near the signal peak (d). It vanishes at later times, once the pulses no longer overlap (e), despite the continued presence of excited-state population.

These results demonstrate that the interference contribution is confined to early times when the two excitation pathways overlap coherently, and is strongly suppressed at later delays. This justifies the use of late-time measurements to isolate photoemission matrix-element effects, as discussed in section~\ref{SM:late_time_matrix_elements}.

\subsection*{Phase reconstruction time dependence}
To maximize the measurement, the phase reconstruction in the main text was performed at the signal peak delay time ($t_1$). While the interference contribution is near its maximum at this time, as presented in Fig.~\ref{fig:SM_CD_TDC}, some additional scattering can be present. These scattering effects are expected to be minimal at zero delay, $t_0$, where the interference contribution is maximum.

Figure \ref{fig:SM_comp_t0_t1} shows the reconstructed phase at both zero delay (panels a-b) and the signal peak (panels c-d). The phase maps for each circular polarization of LE exhibit similar features at both time delays. This is highlighted in their MDCs (panels e-f) where the $t_0$ (gray) curve and the $t_1$ (orange) show minimal differences. The expected difference can be recognized in the extracted relative weight $|\psi_2|^2\cdot|\frac{A}{I_S}|^2\sim 0.014$ ($|\psi_2|^2\cdot|\frac{A}{I_S}|^2\sim 0.012$) at $t_0$ ($t_1$).

\begin{figure}[h]
    \centering
    \includegraphics[width=0.8\linewidth]{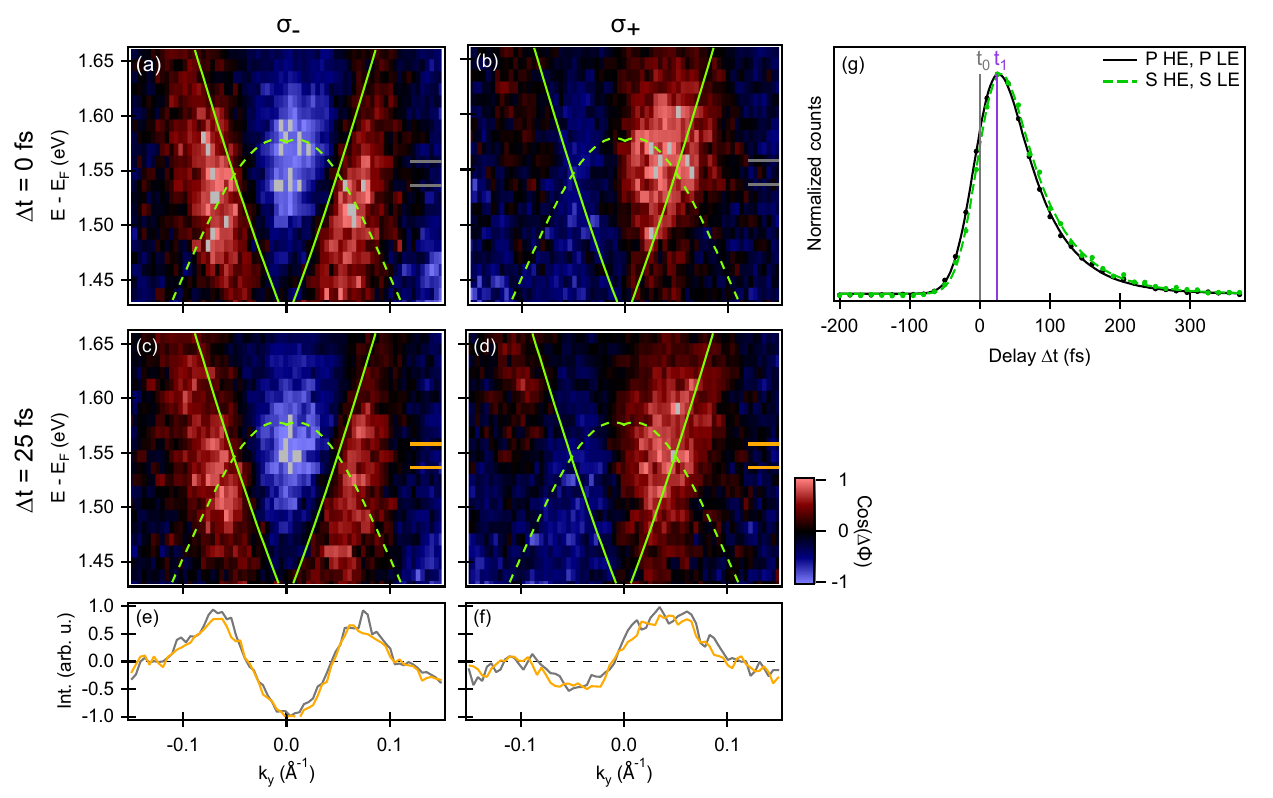}
    \vspace{-5mm}
    \caption{\textbf{Phase reconstruction at $t_0$ and $t_1$}. (a-b) Experimental $\cos(\Delta \phi_-)$ and $\cos(\Delta \phi_+)$ for $\Delta t=0$~fs ($t_0$) respectively. (c-d) Experimental $\cos(\Delta \phi_-)$ and $\cos(\Delta \phi_+)$ for $\Delta t=25$~fs ($t_1$) respectively. (e-f) MDC of $\cos(\Delta \phi_-)$ and $\cos(\Delta\phi_+)$ at $t_0$ ($t_1$) in gray (orange) respectively. (g) Temporal behavior relevant to LE=3.08 eV and HE=5.99 eV, where $t_0$ and $t_1$ are marked in gray and purple, respectively.}
    \label{fig:SM_comp_t0_t1}
\end{figure}

These results were obtained from a dataset in which the LE pulse had a duration of 45~fs, yielding an overall cross-correlation of 60~fs. Therefore, the $t_1=25$~fs time delay in Fig.~\ref{fig:SM_comp_t0_t1} is comparable to the $\Delta t=+70$~fs peak time in the main text results. Importantly, the experimental phase maps in Fig.~4 are similar to those in Fig.~\ref{fig:SM_comp_t0_t1} for each LE polarization at both delay times. Hence, it can be established that the results in the main text are performed within the time frame of the coherent overlap of the two pathways.  

\section{Evaluation of S/P matrix elements at late times}\label{SM:late_time_matrix_elements}

To directly assess the role of photoemission matrix elements in the observed signal, we compare measurements performed with S- and P-polarized HE excitation at late time delays, where interference between the two QPs is strongly suppressed (see section~\ref{SM:time_dependence}).

At sufficiently long pump--probe delays, the coherent overlap between the two excitation pathways is reduced, and the measured signal is dominated by the incoherent population of the excited states. In this regime, any differences between S- and P-polarized HE measurements arise primarily from photoemission matrix elements and final-state effects, without significant contributions from QP interference.

\begin{figure}[h]
\centering\includegraphics[width=0.75\textwidth]{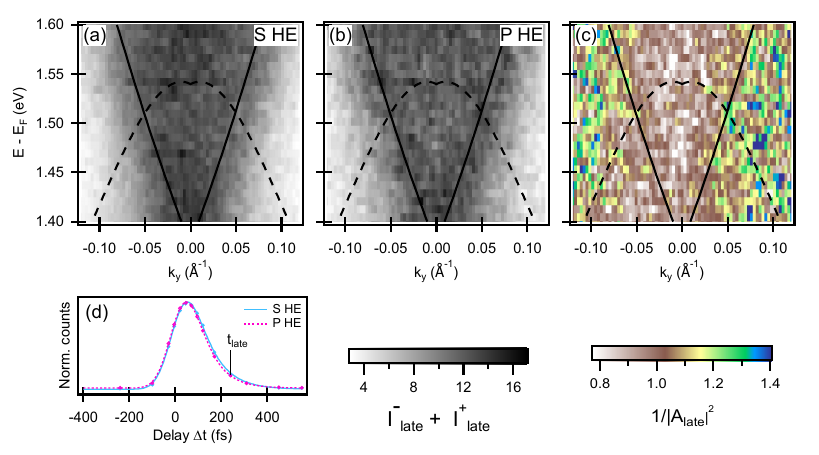}
\vspace{-5mm}
\caption{\textbf{$I_P / I_S$ at later times.} (a-b) $I^- + I^+$ spectrum at $\mathrm{t_{late}}=240$~fs for S-polarized HE and P-polarized HE respectively. (c) $I_P / I_S$ for the late spectrum (a) and (b). (d) Temporal behavior of $I^- + I^+$ for S-polarized (P-polarized) HE in light blue (pink).}
\label{fig:SM_A_late_times}
\end{figure}

Figure~\ref{fig:SM_A_late_times}(a-b) shows the photoemission spectra obtained with S- and P-polarized HE pulses at a delay of $t_{\mathrm{late}} = 240$~fs (d). The spectra are nearly identical, with no significant energy- or momentum-dependent differences in the region of interest. 

To quantify this, we extract the matrix-element ratio $1/|A|^2$ (as defined in Eq.~(3) of the main text) from the late-time data (Fig.~\ref{fig:SM_A_late_times}(c)). This ratio is found to be approximately constant as a function of energy, with only a weak momentum dependence. Importantly, there is no feature that follows the Dirac band dispersion or exhibits a change across $k_y = 0$.

These observations demonstrate that photoemission matrix elements alone do not produce energy- or momentum-dependent features resembling those observed in the phase reconstruction. In particular, they cannot account for the sharp sign change and band-following structure seen in the CD signal at early times. We therefore conclude that the observed phase structure is not explained by photoemission matrix-element effects, and is consistent with the presence of coherent QP interference.

\section{Simulation of circular dichroism with and without quantum path interference}\label{SM:CD_simulation}

In order to identify experimental signatures that distinguish coherent QP interference from matrix-element effects, we consider a minimal model of the 2PPE process and evaluate the resulting circular dichroism (CD) signal under different assumptions. We model the resonant pathway (QP1) as a complex amplitude $\psi_1(E_k)$ with a Lorentzian spectral profile centered at the resonance energy ($E-E_r=0$) and an associated arctangent phase evolution across the resonance, as expected for a driven resonant transition (Fig.~\ref{fig:SM_simulation}(a)). The non-resonant pathway (QP2) is modeled as a weak, energy-independent amplitude $\psi_2$ with constant phase (b). For simplicity, we assume that QP2 carries no intrinsic CD, such that $\psi_2^{+} = \psi_2^{-}\equiv\psi_2$.

\begin{figure}
\centering\includegraphics[width=0.98\linewidth]{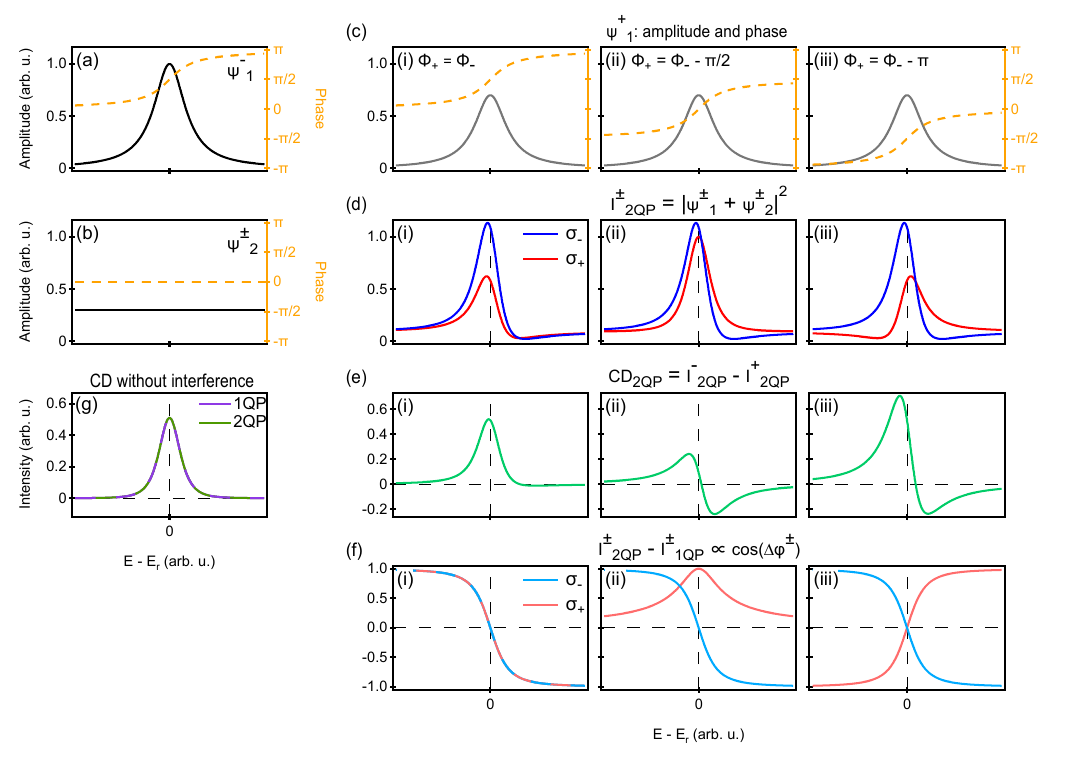}
\vspace{-6mm}
\caption{\textbf{Simulation of CD with and without QP interference.}
(a) Energy-dependent amplitude (solid) and phase (dashed) of a resonant wavefunction $\psi_1^{-}(E_k)$, modeled as a Lorentzian with an arctangent phase evolution across the resonance. (b) Second pathway $\psi_2$, taken as non-resonant with constant amplitude and phase and no intrinsic CD ($\psi_2^-=\psi_2^+$). (c)-(i-iii) Same as (a) for $\psi_1^{+}$, illustrating three relative phase offsets between the $\sigma_+$ and $\sigma_-$ cases: 0, $\pi/2$ and $\pi$ respectively. (d) Resulting intensities for two-quantum-paths (2QP) assuming coherent summation, $I_{2\mathrm{QP}}^\pm = |\psi_1^\pm + \psi_2^\pm|^2$, shown for the three phase scenarios (i-iii), with $\sigma_-$ (blue) and $\sigma_+$ (red). (e) Corresponding CD signal, $\mathrm{CD}_{2\mathrm{QP}} = I_{2\mathrm{QP}}^- - I_{2\mathrm{QP}}^+$. (f) Extraction of $\cos(\Delta\phi_\pm)$ from the simulated data, demonstrating sensitivity to the phase difference between the two paths. (g) Reference CD signal in the absence of interference (single-path in purple and incoherent two-path sum in green), showing a lineshape identical to both and lacking the sign reversal.}
\label{fig:SM_simulation}
\end{figure}

We first consider the case in which only a single pathway contributes (1QP), corresponding to the situation in which QP2 is absent. In this case, the photoemission intensity is given by $I^{\pm}_{\mathrm{1QP}} = |\psi_1^{\pm}|^2$, and the resulting CD signal reflects only differences in the amplitude of the excitation for the two circular polarizations, $\psi^-_1$ (panel a) and $\psi^+_1$ (panels c).
Importantly, while the magnitude of the CD varies across the resonance, its lineshape does not exhibit a sign reversal at the resonance energy. Moreover, since there is no interference here, the result is identical with each of the phase shifts presented in c(i-iii).

Next, we consider a two-path scenario where the contributions of QP1 and QP2 are added \textit{incoherently}, $I^{\pm}_{\mathrm{2QP,inc}} = |\psi_1^{\pm}|^2 + |\psi_2|^2$. In this case, the CD signal (green in panel g) remains qualitatively similar to the single-path result, with only a small modification of its magnitude. In particular, no sign change across the resonance is observed, and there is no dependence on the phase shift between $\psi_1^+$ and $\psi_1^-$.

In contrast, when the two pathways are added coherently, $I^{\pm}_{\mathrm{2QP}} = |\psi_1^{\pm} + \psi_2|^2$ (Fig.~\ref{fig:SM_simulation}(d) i-iii), the resulting CD signal exhibits a qualitatively distinct behavior (Fig.~\ref{fig:SM_simulation}(e) i-iii).
The three sub-panels represent different phase shifts between the $\sigma_+$ and $\sigma_-$ cases (as plotted in panels c(i-iii)). Notably, this phase shift is now imprinted on the intensities and on the CDs, though it does not influence either the 1QP case nor the incoherent case (Fig.~\ref{fig:SM_simulation}(g)).
Due to the $\pi$-phase shift acquired by the resonant pathway ($\psi_1^{\pm}$) across the resonance, the interference term changes sign, leading to a sharp sign reversal of the CD at (or near) the resonance energy. This behavior is robust for a range of non-zero relative phases between the two circular polarizations, including the experimentally relevant case (ii) of a relative phase shift close to $\pi/2$, though at phase shift of zero there is no sign change.

Furthermore, the simulated $\cos(\Delta\phi_\pm)=(I_\mathrm{2QP}^\pm-I_\mathrm{1QP}^\pm)/(2\sqrt{I_\mathrm{1QP}^\pm}\cdot |\psi_2^\pm|)$, is presented in Fig.~\ref{fig:SM_simulation}(f), with a good agreement between experiment and the $\pi/2$ case (ii). The $k$-dependence of the phase is, of course, not reflected in this simulation, which assumes a generic resonance at a single EDC. 

While both the single-path and incoherent two-path cases yield CD signals without sign reversal, only the coherent two-path scenario reproduces a sign change across the resonance (panels (e)i-iii). This establishes that such a feature is a direct fingerprint of coherent interference between quantum pathways. 
Importantly, this conclusion does not rely on any assumptions regarding the similarity of photoemission matrix elements for different polarizations or final states, as it is obtained within a single HE polarization configuration. The presence of a sign change in the CD signal therefore provides direct experimental evidence for QP interference, independent of matrix-element effects. Additionally, it reveals the existence of a phase difference between $\psi_1^-$ and $\psi_1^+$.

\section{Phase reconstruction with linear low-energy polarization}
\label{SM:linear_LE}
In the main text, the phase reconstruction is carried out using circularly polarized LE light, motivated by the strong dichroic response of the Dirac cone. An alternative explanation for the observed signal could be that it arises from photoemission-CD associated with the second pathway (QP2), rather than from interference between the two pathways. While modifications to the spectrum due to scattering states in CD-ARPES have been observed, they are expected to be weak at low-photon energies such as LE and HE used in this work\cite{boban_scattering_2025}. Nevertheless, any signal of this origin would be expected to vanish when using linearly polarized LE.

\begin{figure}[h]
\centering\includegraphics[width=0.7\linewidth]{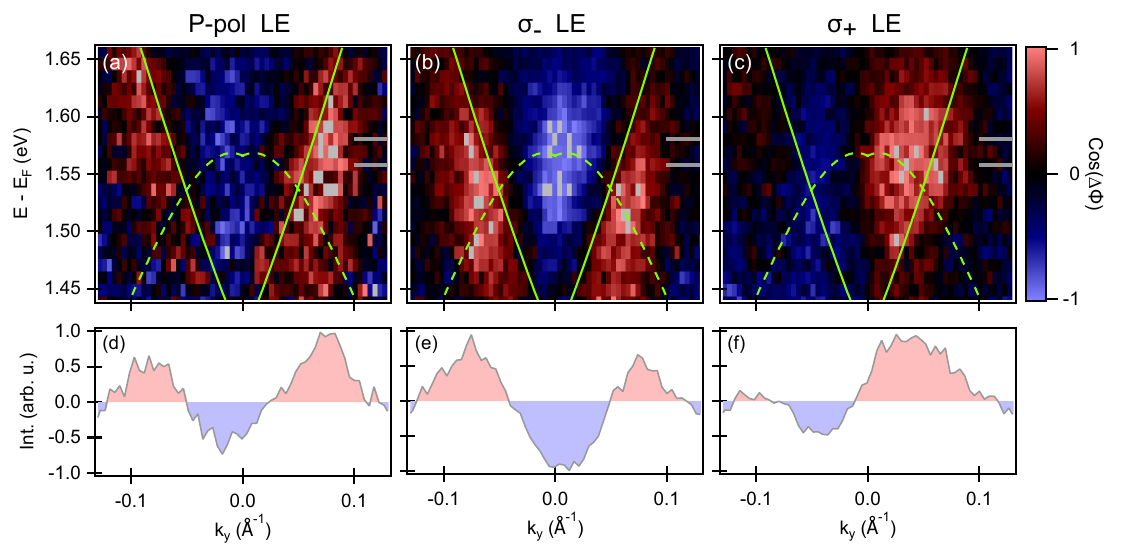}
\vspace{-5mm}
\caption{\textbf{Phase reconstruction with linear vs circular LE polarization.} (a-c) Experimental $\cos(\Delta\phi_{P})$, $\cos(\Delta\phi_-)$ and $\cos(\Delta\phi_+)$ respectively, measured at $\Delta t=0$~fs with LE=3.08~eV and HE=5.99~eV. (d-f) Momentum Distribution Curves (MDCs) of $\cos(\Delta\phi_{P})$, $\cos(\Delta\phi_-)$ and $\cos(\Delta\phi_+)$ respectively for the energy range between the gray lines.}
\label{fig:SM_phase_P_pol}
\end{figure}

Figure~\ref{fig:SM_phase_P_pol} presents the reconstructed $\cos(\Delta\phi)$ obtained using linearly (P) polarized LE excitation, alongside the corresponding results for $\sigma_{-}$ and $\sigma_{+}$ polarization. The phase map of the P-polarized LE case and its momentum distribution curve (MDC) exhibit the same qualitative features as in the $\sigma_-$ case. In particular, the resonance-induced phase variation across the Dirac band and the position of the zero-crossing are preserved.

The results presented in Fig.~\ref{fig:SM_phase_P_pol} are from a different data set than the ones in the main text. They were performed with a HE pulse of 5.99 eV, a cross-correlation time resolution of $\sim70$~fs and at delay $\Delta t=0$~fs. 

The persistence of the phase structure with linear LE polarization demonstrates that the observed signal is not due to CD of QP2, but instead originates from the resonant intermediate state and reflects the properties of the Dirac cone.

\section{Sensitivity of phase reconstruction to QP2 (IPS-mediated pathway)}\label{SM:QP2_sensitivity}

\begin{figure}
\centering\includegraphics[width=0.75\linewidth]{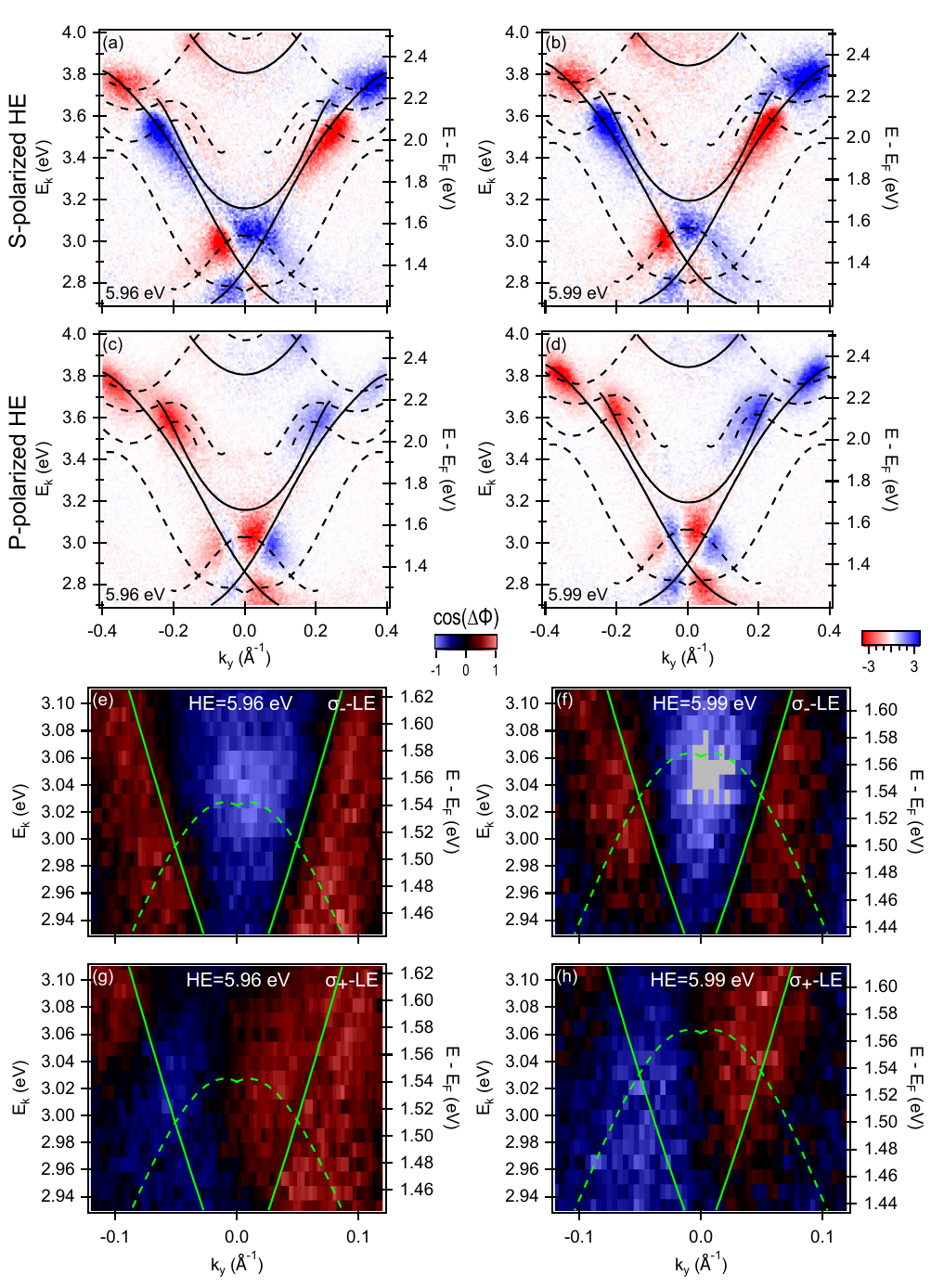}
\vspace{-8mm}
\caption{\textbf{CD and phase reconstruction with HE pulse energy shifted by 30 meV.} (a-b) $I^- - I^+$ measured for S-pol HE with HE=5.96~eV and HE=5.99~eV, respectively. (c-d) $I^- - I^+$ measured for P-pol HE with HE=5.96~eV and HE=5.99~eV, respectively. (e-f) Reconstructed $\cos(\Delta \phi_-)$ with HE=5.96~eV and HE=5.99~eV, respectively. (g-h) Reconstructed $\cos(\Delta \phi_+)$ with HE=5.96~eV and HE=5.99~eV, respectively. The HE=5.96~eV spectra are the same as in the main text, with (a,c) the expanded range of Fig.~3(c,d) and (e,g) the same as Fig.~4(a,b).
The results with HE=5.99~eV are shown at $\Delta t=25$~fs when the resonance signal is maximal.}
\label{fig:SM_different6eV}
\end{figure}

The phase reconstruction relies on interference between a resonant pathway (QP1) and a non-resonant reference pathway (QP2), mediated by the image potential state (IPS). The influence of QP2 on the reconstructed phase is therefore determined by its detuning from the IPS resonance and its spectral weight.

To probe this dependence, we compare measurements acquired with HE photon energies of 5.96~eV and 5.99~eV (Fig.~\ref{fig:SM_different6eV}), corresponding to a shift of the resonance position by $\sim 30$~meV and a change in IPS detuning of $\sim 15$~meV. The CD patterns and reconstructed phase maps are unchanged within experimental resolution. In particular, the phase zero-crossing and the sign change across the Dirac band remain fixed, indicating that the extracted phase is insensitive to moderate variations in QP2 detuning. Across all datasets measured in our setup, variations in photon energy and sample doping shift the relative IPS position by several tens of meV, yet the characteristic Fano-like CD lineshape and interference signatures remain identical. These observations show that the interference structure is not controlled by the precise energetic position or coupling strength of the IPS.

Within the reconstruction procedure, QP2 enters through its intensity contribution $|\psi_2|^2$, which affects only the normalization in Eq.~(3) of the main text. The energy- and momentum-dependent phase structure is determined by the interference term. Varying $|\psi_2|^2$ over a broad range, including replacing the Lorentzian IPS lineshape with a constant value, does not modify the reconstructed phase pattern.

The key experimental signature -- the sharp CD sign change across the band maximum -- is already present in the raw data for P-polarized HE excitation. As shown in section~\ref{SM:CD_simulation} (Fig.~\ref{fig:SM_simulation}), this feature is a generic consequence of coherent interference between a resonant and a non-resonant pathway, and does not rely on specific assumptions about the IPS.

These results establish that the reconstructed phase is robust with respect to the properties of QP2, and that its structure is governed by the resonant pathway (QP1).

\end{supp}

\end{document}